\documentclass[11pt,a4paper]{article}
\usepackage[german, english]{babel}
\usepackage{a4wide}
\usepackage{amsmath}
\usepackage{amssymb}
\usepackage{ifthen}
\usepackage{epsfig}
\newcounter{fig}   \newcommand{\lbfig}[1]{\refstepcounter{fig}
\label{#1} } 

\newcommand{\hmu}{\hat\mu}
\newcommand{\hnu}{\hat\nu}

\renewcommand{\det}{\operatorname{Det}}

\newcommand{\Gr}[2][]{\ensuremath{ \mathrm{#2}
\ifthenelse{\equal{#1}{}} {} {(#1)} } } 
\newcommand{\lie}{\mathfrak}
\newcommand{\gr}[2][]{\ensuremath{\lie{#2} \ifthenelse{\equal{#1}{}}
{} {(#1)} } }     
\newcommand{\minuspt}{\\[-4pt]}
\newcommand{\abb}[3]{\ensuremath{ \ifthenelse{\equal{#1}{}}{}{#1:} #2
\mapsto #3}} 
\newcommand{\abbildung}[6][]{\begin{align}
\ifthenelse{\equal{#1}{}}{}{\label{#1}}
\ifthenelse{\equal{#2}{}}{}{#2:} #3 & \rightarrow   #4  \minuspt #5 &
\mapsto #6  \notag \end{align}}

\def\Det{\mathop{\rm Det}\nolimits}
\def\gl22{\mathop{\mathfrak{gl}(2|2)}\nolimits}

\def\O{\mathcal{O}}

\def\Tr{{\rm Tr}}

\def\sq3{\sqrt{3}}
\def\isq3{\frac{1}{\sqrt{3}}}

\def\defi{\stackrel{\rm def}{=}}
\newcommand\grsim{\mathrel{\hbox{\lower1ex\hbox{\rlap{$\sim$}\raise1ex\hbox{$>$}}}}}
\newcommand\losim{\mathrel{\hbox{\lower1ex\hbox{\rlap{$\sim$}\raise1ex\hbox{$<$}}}}}

\newcommand{\npmuh}{\ensuremath{({\scriptstyle n+\frac{\hat\mu}{2}})}}
\def\npth{\ensuremath{({\scriptstyle n+\frac{\hat{2}}{2}})}}
\def\npoh{\ensuremath{({\scriptstyle n+\frac{\hat{1}}{2}})}}
\def\npthh{\ensuremath{({\scriptstyle n+\frac{\hat{3}}{2}})}}
\def\nmoh{\ensuremath{({\scriptstyle n-\frac{\hat{1}}{2}})}}
\def\nmth{\ensuremath{({\scriptstyle n-\frac{\hat{2}}{2}})}}
\def\nmthh{\ensuremath{({\scriptstyle n-\frac{\hat{3}}{2}})}}

\newcommand{\nmmuh}{\ensuremath{({\scriptstyle n-\frac{\hat\mu}{2}})}}
\newcommand{\npnuh}{\ensuremath{({\scriptstyle n+\frac{\hat\nu}{2}})}}
\newcommand{\nmnuh}{\ensuremath{({\scriptstyle n-\frac{\hat\nu}{2}})}}

\newcommand{\npmunuh}{\ensuremath{({\scriptstyle n+\hat\mu+\frac{\hat\nu}{2}})}}
\newcommand{\npnumuh}{\ensuremath{({\scriptstyle n+\hat\nu+\frac{\hat\mu}{2}})}}
\newcommand{\nprhoh}{\ensuremath{({\scriptstyle n+\frac{\hat\rho}{2}})}}
\newcommand{\nmrhoh}{\ensuremath{({\scriptstyle n-\frac{\hat\rho}{2}})}}
\newcommand{\npetah}{\ensuremath{({\scriptstyle n+\frac{\hat\eta}{2}})}}
\newcommand{\nmetah}{\ensuremath{({\scriptstyle n-\frac{\hat\eta}{2}})}}
\newcommand{\npalh}{\ensuremath{({\scriptstyle n+\frac{\hat\alpha}{2}})}}
\newcommand{\nmalh}{\ensuremath{({\scriptstyle n-\frac{\hat\alpha}{2}})}}
\begin{document}

\begin{titlepage}

  \title{The Color--Flavor Transformation of induced QCD}

\author{
Yasha Shnir
\\ [2mm]
{\it {\footnotesize Institute of Physics, Carl von Ossietzky University Oldenburg}}\\
{\it {\footnotesize  D 26111 Oldenburg, Germany}}
\\ [2mm]
}

\date{~}

\maketitle 
\begin{abstract}
  The color--flavor transformation is applied to the $U(N_c)$ 
lattice gauge model, in which the gauge theory is induced by a heavy chiral 
scalar field sitting on lattice sites. The flavor degrees of 
freedom can encompass several `generations' of the auxiliary
field, and for each generation, remaining indices are associated 
with the elementary plaquettes touching the lattice site. 
The effective, color-flavor transformed theory 
is expressed in terms of gauge singlet matrix fields carried by lattice 
links.  
The effective action is analyzed for a hypercubic lattice 
in arbitrary dimension. 
The saddle points equations
of the model in the large-$N_c$ limit are discussed. 

\end{abstract}

\bigskip



\end{titlepage}


\section{Introduction}
The complexity of quantum chromodynamics (QCD) originates from the random 
character of the gauge field in the low-energy regime, while at high 
energy (small scales) the theory is asymptotically free. 
One of the most successful approaches to analyse QCD in the non-perturbative 
domain is   
the lattice formulation due to Wilson 
\cite{Wilson74}, where the strong coupling regime becomes natural. 
On the other hand, the continuum limit of lattice QCD corresponds to the 
weakly-coupled regime. It was shown by Gross and Witten \cite{Gross80} 
that two-dimensional lattice QCD can be solved exactly in the large-$N_c$ 
limit and there is a third-order  weak- to strong-coupling 
phase transition. The problem, which persists over the years is, 
if such a phase transition occurs in the 
realistic $SU(3)$ four-dimensional gauge theory. 

One of the ways to attack this problem is to use some form of 
\emph{duality} \cite{Banks77} which appears in the lattice theory.  
The general idea of duality is that a given theory
can have two, or more equivalent formulations, with   
different sets of fundamental variables. 
Usually these dual
formulations are related by interchanging a parameter, e.g. the 
electromagnetic coupling
constant $e^2$, with its inverse $1/e^2$. 
For example, 
in the weak-coupling limit, the action of the Abelian lattice 
gauge model can be approximated by the Villain form 
(see e.g. \cite{Banks77,Peskin78}) which allows to define the 
dual variables. Similarly, it was shown that there is 
a duality transformation from the compact $U(1)$
gauge theory into a non-compact Abelian Higgs model \cite{Polley91}.

The dual variables are in general not defined on the
original lattice but on a \emph{dual lattice}. For a hypercubic lattice, 
it is obtained by shifting the 
original lattice by half of the lattice spacing in all dimensions. 
Thus, the lattice duality not only transforms  the 
variables of the functional integral, but also incorporates a
transfer to the dual lattice. 

In the case of a nonabelian lattice gauge theory, one can 
reformulate the model in terms 
of plaquette variables \cite{Halpern79,Rusakov97}.
Recently, following the idea of \cite{Anishety93},
Diakonov and Petrov \cite{DP2000} applied a
Fourier transformation to write down the Jacobian from the link variables 
to the plaquette variables in $d=3$ gluodynamics for the gauge group
SU$(2)$;
\emph{in fine} the dual lattice is composed of tetrahedra representing
$6j$-symbols, with links of arbitrary lengths.
In the continuum limit this effective theory is equivalent 
with quantum gravity with the Einstein-Hilbert action.  
Unfortunately, this scheme seems difficult to apply to higher-rank gauge groups
and in $d > 3$ dimensions.

Another approach to analyze non-perturbative QCD, the so-called 
``Induced QCD'', was initiated in the 90s \cite{Kazakov93,Makeenko93}. 
This direction originated from A.D.~Sakharov's idea to 
treat the gauge theory as induced quantum field theory.
In the Kazakov-Migdal 
model \cite{Kazakov93} the Wilson lattice action is induced by the 
auxiliary heavy 
scalar matrix fields which are taken in an 
adjoint representation of the gauge group $SU(N_c)$.  
This field can be diagonalized by a gauge transformation so that in the 
large $N_c$ limit the functional integral over eigenvalues of the matrix 
field, which serves as a master field,
is saturated by a saddle point of the effective action. The induced action
obtained via integration over the auxiliary fields contains the traces 
of products of the link variables along all possible contours. However in the
large mass limit, only one-plaquette loops survive and the Wilson action 
is recovered.   

The interest in the Kazakov-Migdal
model was mainly inspired by its exact solvability 
in the large $N_c$ limit but its continuum 
limit is questionable. Moreover, for inducing fields in adjoint representation 
there is an extra local $Z_N$ symmetry which leads 
to infinite string tension (the so-called local confinement) 
rather than the conventional
area law for the Wilson loop \cite{Kogan92}. That is also the case for the 
adjoint fermion model of the induced QCD  \cite{Makeenko93}.

Another example of the QCD inducing model was proposed by Bander and Hamber  
\cite{Bander,Hamber}. In this approach the Wilson action is recovered 
if the number of flavors of the auxiliary fields 
goes to infinity simultaneously with the mass. This model 
does not suffer from the extra gauge symmetry \cite{Aref,Zarembo} but 
is not solvable even in $d=2$.
 
In the present note, based on a work in collaboration with 
S.~Nonnenmacher \cite{ShNonn}, I discuss a different approach to treat a 
similar type of inducing model. Our construction 
starts from an inducing theory similar with
\cite{Bander,Hamber}, already introduced in \cite{J-diss,ShNonn,SchWett}, 
and applies a certain duality transformation, namely
the ``color--flavor transformation'' \cite{Zirn}. 
We have recently applied 
this transformation to the lattice SU$(N_c)$ model in the strong-coupling
limit, which describes  
quarks coupled with a background gauge field \cite{BNSZ}, see also 
paper \cite{SchWett}. 
Schlittgen and Wettig also 
independently applied the SU$(N)$ color-flavor transformation to a 
similar, yet different QCD-inducing 
model  \cite{su-rivals,SchWei}. 
A very interesting formalism to induce lattice gauge model is suggested by 
Budczies and Zirnbauer in paper \cite{BuZirn}, 
where instead of heavy inducing fields a finite number $N_b$ of auxiliary boson flavors
was coupled to the gauge field. In this framework $U(N_c)$ gauge theory is induced 
when $N_b$ exceeds $N_c$ and the boson mass is lowered to a critical point. 
In the present note I would like to sum up our investigation of a 
possibility to apply another  ``color--flavor motivated'' 
approach, which is related with some modification of the model by Bander and Hamber. 


\section{A model of induced lattice gauge theory}
\subsection{Wilson's lattice action}
We consider a Euclidean U$(N_c)$ pure gauge action (no quarks)
in $d$ dimensions, 
placed on a hypercubic lattice with lattice constant $a$. The choice 
of U$(N_c)$ instead of the realistic SU$(N_c)$ highly simplifies the 
subsequent color-flavor transformation. 
 
The lattice sites are labeled by integer vectors $n =
(1, \ldots, n_d)$, 
the gauge matrix variables 
\begin{equation}   \label{link-matrices}
U_\mu(n) \equiv U_{n, n+\mu} \equiv
U \npmuh = \exp
\left( iag A_\mu(na + \frac{a\hat\mu}{2}) \right) \in {\rm U}(N_c)
\end{equation}
are placed on the lattice links $n + \hat\mu/2$ (we label links by their
\emph{middle points}), 
leaving the site $n$ in any of the ``positive''
directions $\mu = 1,\ldots,d$. The plaquettes are either labeled by 
an independent index $p$, or by triplets of the form $(n,\pm\mu,\pm\nu)$. 
For instance, the 
plaquette $(n,\mu,\nu)$ contains the links $n + \hat\mu/2$ and 
$n + \hat\nu/2$. To fix an
ordering between the directions, we will in general assume that 
$1\leq\mu<\nu\leq d$. 
Notice that the same plaquette corresponds to the triplets $(n,\mu,\nu)$ and
$(n+\hat\mu,-\mu,\nu)$ (as well as two other triplets).

The Wilson pure gauge action is given by a sum over all elementary plaquettes:
\begin{equation}     
\label{action}
-S_{\rm gluons} = \beta_W\sum\limits_{p} 
\Tr \left(U_{P}(p) + U_{P}^\dagger(p)\right).
\end{equation}
Here $\beta_W$ is the lattice coupling, which is related to the
bare continuum coupling constant $g$ through
\begin{equation} \label{beta-g}
\beta_W=\frac{a^{d-4}}{2g^2}.
\end{equation}
The plaquette field $U_P$ 
is defined as an ordered product 
of the link variables along the boundary of the given plaquette:
\begin{equation} \label{plaquette}
U_P(n,\mu,\nu) =  U \npmuh U\npmunuh U^{-1} \npnumuh U^{-1}\npnuh  \, . 
\end{equation}
The partition function is defined as 
\begin{equation}    \label{partition}
Z = \int {\cal D} U e^{-S_{\rm gluons}},
\end{equation}
the invariant measure of integration is defined as a product over all links
${\cal D} U = \prod\limits_{n,\mu} dU\npmuh $ and $  dU\npmuh $ is the Haar 
measure on the group U$(N_c)$.

Using a generalized Baker-Campbell-Hausdorff formula, one can relate
the continuum field strength tensor 
$$
F_{\mu\nu} = \partial_\mu A_\nu - \partial_\nu A_\mu - i[A_\mu, A_\nu]
$$
with the plaquette matrices as follows:
\begin{equation}  \label{gauge-tensor}
U_P(n,\mu,\nu) = e^{iga^2 F_{\mu\nu}(n) + O(a^3)} \, .
\end{equation} 
Expansion in the lattice spacing up to the second order then yields
$$
\Tr~ U_P(n,\mu,\nu) \approx N_c + iga^2~ \Tr~ F_{\mu\nu} - \frac{g^2a^4}{2}
 ~\Tr~ F_{\mu\nu}^2 \, ,
$$
so that the standard Yang-Mills action is 
recovered in the continuum limit:
\begin{equation}   \label{cont-action}
S_{cont} = \frac{1}{2} \int d^d x~ \Tr~ F_{\mu\nu}^2.
\end{equation}

Wilson's lattice gauge action (\ref{action}) is written via the plaquette 
matrices $U_P(p)$ while the variables of the integration measure
are the link matrices $U\npmuh$. 
It is possible to explicitly transfer the integration from link to 
plaquette matrices \cite{Halpern79,Rusakov97,DP2000} , but this procedure 
is technically involved and  
can be applied only in a few simple cases. 
We would like to investigate a possibility to apply another approach, 
which starts from a modification 
of the QCD inducing model of Bander and Hamber  
\cite{Bander,Hamber}. 


\subsection{Description of the model and its induced gauge action}
Let us consider a massive complex bosonic field $\phi(n)$ placed  
on the lattice sites $n$. This field has ``flavor''components 
$\phi^{(\pm\mu,\pm\nu)}(n)$ associated to each of the $2d(d-1)$
plaquette $\{(n,\pm\mu,\pm\nu);1\leq\mu<\nu\leq d\}$ adjacent to the site $n$
 (see Figures~\ref{def-fields},\ref{ind-link}).
The field furthermore decomposes into
two ``chiral components'' 
$\phi_{R}^{(\mu,\nu)}(n)$ and $\phi_{L}^{(\mu,\nu)}(n)$,
which are hopping in opposite directions. These fields will be referred  
to as the ``left'' component and the ``right'' component  respectively. 

The chiral bosonic field can be thought of 
as an $N_b$-component vector in an auxiliary 
``flavor'' space (here the index $b$ stands for ``bosonic''). The number of 
``flavors'' has to be a multiple of  $2d(d-1)$, that is  
the dimension of the ``flavor'' space is $N_b=n_b \times 2d(d-1)$, where 
$n_b \in  \mathbb{N}^*$ is the number of ``generations'' of the 
bosonic field. The bosonic field $\phi$ also 
transforms as a vector through 
the gauge group $U(N_c)$, so it contains ``color indices''   
$i =1,\dots, N_c $ besides the ``flavor'' indices $a = 1,\dots, N_b$. All
flavor components have the same mass $m_b$.

\begin{figure}[t]
\begin{center}
  \setlength{\unitlength}{1cm}
\begin{picture}(13,7)
  \put(1.0,0.0) {\mbox{\epsfysize=8.0cm\epsffile{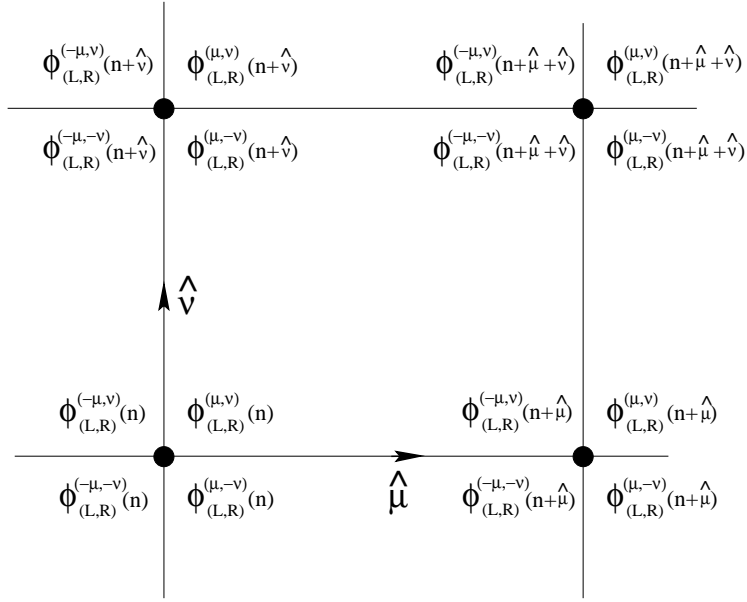}}}
  \lbfig{def-fields}
\end{picture}
\caption{Auxiliary chiral fields in the plane $(\mu\nu)$. Each field component
is written inside the plaquette it is associated with.}
\end{center}
\end{figure}

This model (already considered in \cite[Chap. 5]{J-diss} and \cite{ShNonn}) is
different from the model usually considered in induced QCD
\cite{Bander,Hamber,Suranyi}, where the flavor degrees of freedom of the 
auxiliary fields are not associated with plaquettes. As we will show below, 
this structure will induce a Wilson-type action in a cleaner way than
in the previous models.

To complete our notations, we shall fix the orientation of the plaquettes, 
in order to define the hopping and the ``left'' and the ``right''  
components 
in all $2$-dimensional planes on the lattice. On any plane
$(\mu,\nu)$ with $1\leq\mu<\nu\leq d$
the `left' chiral component hops on the plaquette $(n,\mu,\nu)$ as
\begin{equation}        \label{left-convention}
\phi_L (n) \to \phi_L(n+\hmu) \to \phi_L(n+\hmu +\hnu)
\to \phi_L(n +\hnu) \to \phi_L(n),
\end{equation}
and the `right' chiral component hops in the opposite direction 
(see Fig.~\ref{ind-plaq}). This practically means that the ``kinetic''
part of the action contains the term (cf. Eq.~\eqref{action-boson-1}): 
$$
-\phi^{(-\mu,\nu)\dagger}_L(n+\hmu)U\npmuh \phi^{(\mu,\nu)}_L (n) \, .
$$

Let us now group
the fields surrounding a given plaquette
$p=(n,\mu,\nu)$ into the following \emph{plaquette quadruplets}:
\begin{equation}         \label{boson-multiplets}
{\phi}_{L}(p) \defi 
\begin{pmatrix} 
{\phi}^{(\mu,\nu)}_{L}(n)\\ 
{\phi}^{(-\mu,\nu)}_{L}(n+\hmu)\\
{\phi}^{(-\mu,-\nu)}_{L}(n+\hmu+\hnu)\\
{\phi}^{(\mu,-\nu)}_{L}(n+\hnu)
\end{pmatrix};\qquad 
{\phi}_{R}(p) \defi 
\begin{pmatrix} 
{\phi}^{(\mu,\nu)}_{R}(n)\\ 
{\phi}^{(-\mu,\nu)}_{R}(n+\hmu)\\
{\phi}^{(-\mu,-\nu)}_{R}(n+\hmu+\hnu)\\
{\phi}^{(\mu,-\nu)}_{R}(n+\hnu)
\end{pmatrix}
\end{equation}
Then the plaquette action of the ``left'' bosonic massive
field may be written in the concise form 

\begin{figure}[t]
\begin{center}
  \setlength{\unitlength}{1cm}
\begin{picture}(13,7.5)
  \put(-1.5,0.5) {\mbox{\epsfysize=7.0cm\epsffile{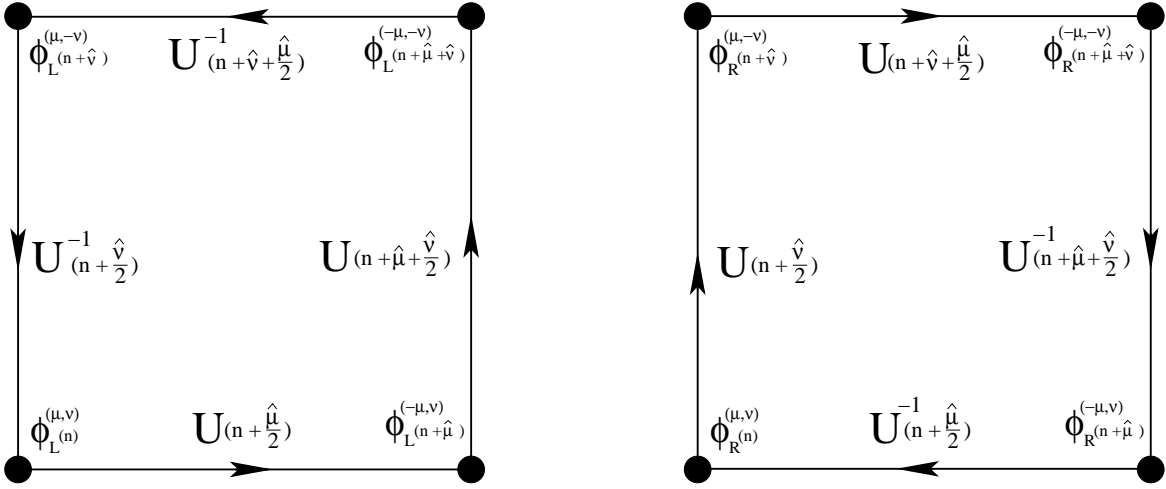}}}
  \lbfig{ind-plaq}
\end{picture}
\caption{Coupling of the ``left'' and ``right'' chiral components of the bosonic 
field with the gauge field along the elementary plaquette $(n,\mu,\nu)$.}
\end{center}
\end{figure}

\begin{equation}   \label{action-boson-1}
S_L(p) = 
\phi^\dagger_{L}(p) M_L(p) \phi_{L}(p) 
\end{equation}    
with the $4N_c\times 4N_c$ matrix
\begin{equation}        \label{M_L}
M_L(n,\mu,\nu) \defi \left( \begin{array}{cccc} 
  m_b     & 0   & 0 & -U^{-1}\npnuh\\ 
-U\npmuh& m_b & 0 & 0 \\
 0        & -U\npmunuh &m_b & 0\\
  0       & 0   & -U^{-1}\npnumuh & m_b
\end{array}\right)
\end{equation}
Similarly, ``right'' bosonic action 
associated with the plaquette $p$ reads:
\begin{equation}   \label{action-boson-2}
S_R(p) = \phi^\dagger_{R}(p) M_L^\dagger(p) \phi_{R}(p). 
\end{equation} 
We define the full partition function as: 
\begin{equation}      \label{z-boson}
 \mathcal{Z} = \int {\cal D}{\bar \phi}_{L,R} {\cal D} \phi_{L,R} {\cal D}U
\ \exp\left\{-\sum_p S_L(p) + S_R(p)\right\}.
\end{equation}


\subsection{Integration over auxiliary fields}
We now show that the model (\ref{z-boson}) 
induces a lattice gluodynamics which reduces to Wilson's pure gauge action for
suitably chosen parameters. We can  treat plaquettes separately, since
each field component is associated with one and only plaquette.
The integration over
the `left' auxiliary fields in (\ref{action-boson-1}) yields 
\begin{equation}
\int {d}{\bar\phi}_L(p) {d} \phi_L(p)\exp\{-S_L(p)\}  = 
\det\left(M_L(p) \right)^{-n_b} 
=\det\left(m_b^4 - U_P(p)\right)^{-n_b},
\end{equation}
where $n_b$ is the number of ``generations'' of the auxiliary fields. The
integration over the ``right'' components is similar, with $U_P$ replaced
by $U_P^\dagger$.
Thus, the integration over the auxiliary
bosonic fields exactly yields the pure gauge effective action
\begin{equation}  \label{induced-act}
\begin{split}
S_{\rm plaq} &= n_b \sum\limits_p [\ln \det(1 - \beta_b  U_P(p)) +  
\ln \det(1 - \beta_b  U_P^\dagger(p))]\nonumber\\
&= n_b \sum\limits_p \Tr [\ln (1 - \beta_b  U_P(p)) +  
\ln (1 - \beta_b  U_P^\dagger(p))],
\end{split}
\end{equation}
where we skipped a mass-dependent prefactor, and set $\beta_b = m_b^{-4}$.
As opposed to the former inducing models \cite{Bander,Hamber}, 
this action does not contain 
any term related with larger loops.
Expanding this action for small parameter $\beta_b$ (that is, large $m_b$), 
we get
\begin{equation}\label{expansion}
S_{\rm plaq} = -n_b \beta_b  \Tr \sum\limits_{p}\left(
U_{P}(p) + U_{P}^\dagger(p)\right)+\O(n_b\beta_b^2).
\end{equation}
This coincides with Wilson's action (\ref{action}) if we identify 
\begin{equation}   \label{g-beta}
\beta_W=n_b\beta_b\Longleftrightarrow g^2 =  \frac{a^{d-4}}{2n_b\beta_b}
=  \frac{a^{d-4}m_b^4}{2n_b}.
\end{equation}


\paragraph{Remarks on the continuum limit}
Let us say a few words about the continuum limit of the model. 
In $d=2$ and $d=3$,
the physical coupling constant $g$ can remain fixed as one lets the 
lattice spacing 
$a$ go to $0$ and simultaneously $\beta_W=n_b\beta_b \to \infty$. 
In $d=4$, this 
limit corresponds to the asymptotically free continuum theory. 
Recall that the mass of the auxiliary fields is measured in units 
of the lattice spacing: $m_b = m\times a$. Therefore 
the auxiliary field becomes non-observable in the continuum limit 
if the corresponding correlation length $\xi = (am)^{-1} = \beta_b^{1/4}$
stays finite. We assumes it remains small enough to justify 
the expansion \eqref{expansion}. 

\medskip

We end this section with one more comment.
It is possible to consider a fermionic counterpart
of the bosonic action \eqref{z-boson}, which induces a similar pure gauge 
effective
action. One has to replace the bosonic auxiliary fields by fermionic
(anticommuting) fields $\psi_{L,R}$, $\bar\psi_{L,R}$ which carry
color indices and flavor indices related to plaquettes, exactly as
for the multiplets \eqref{boson-multiplets}; one can consider 
$n_f$ ``generations'' of these fermions.
After integrating over them, one obtains the effective pure gauge action
\begin{equation}  \label{induced-act-fermions}
-S = n_f \sum\limits_p \Tr~ [\ln (1 + \beta_f  U_P(p)) +  
\ln (1 + \beta_f  U_P^\dagger(p))].
\end{equation}
Here $\beta_f = m_f^{-4}$, where $m_f$ is the fermion mass. 
Clearly, we once more recover
the conventional Wilson action (\ref{action}) for small values of $\beta_f$. 

Having considered both bosonic and fermionic induced lattice 
gluodynamics, we can also represent the effective action as the ratio
of  ``fermionic'' and ``bosonic'' determinants:
\begin{equation}   
\exp \biggl[ \frac{a^{d-4}}{2g^2} \Tr( U_P +  U_P^\dagger)\biggr]
\approx 
\frac{\left[\det(1+\beta_f U_P)\det (1+\beta_f U_P^\dagger)\right]^{n_f}}
{\left[\det(1-\beta_b U_P)\det(1-\beta_b U_P^\dagger)\right]^{n_b}}.
\end{equation} 
Thus, for small couplings $\beta_b$, $\beta_f$, the partition function can be 
represented by the following superintegral 
\begin{equation}
 \mathcal{Z} = \int {\cal D} U \int {\cal D}\psi 
{\cal D} {\bar \psi} \exp \biggl[ 
 -{\bar\psi}^i_{L,a}(\delta^{ij}
+ \beta_{L} (U_P)^{ij}) \psi^j_{L,a}
-{\bar \psi}^i_{R,a} (\delta^{ij}
+ \beta_{R} (U_P^\dagger)^{ij})
\psi^j_{R,a}\biggr].
\end{equation}
The composite field $\psi,\ \bar\psi$ includes both bosonic and fermionic 
variables, which are distinguished by the ``flavor'' index $a$. 

There are therefore several ways to induce Wilson's lattice gauge action. 
In all cases, the action (\ref{action}) with fixed $\beta_W$
can be recovered in the limit of large mass and large number
of generations of the auxiliary fields (cf. Eq.~\eqref{g-beta}). 
Below we will restrict our considerations
to the bosonic model (\ref{z-boson}).  


\section{Color-flavor transformation of the inducing theory}

Though the equivalence between the model (\ref{z-boson}) and Wilson's
gluodynamics can be established only in the limit of large 
number of generations of the auxiliary field, we would like to 
study some properties of the underlying theory with a single bosonic
``generation'' $(n_b=1)$, that is with a flavor space of dimension 
$N_b = 2d(d-1)$.
This assumption simplifies the application
of the color-flavor transformation \cite{Zirn}.

Let us consider the interaction term of the 
bosonic action (\ref{z-boson}) on a given 
link  $(n + \hmu/2)$ of the $d$-dimensional 
hypercubic lattice. The path ordered product of the link matrices 
 defined by the ``left'' action (\ref{action-boson-1}) is depicted in 
Fig.~\ref{ind-link}. 
There are $2d-2$ plaquettes which share this common link.
Above we have used the plaquette quadruplets (\ref{boson-multiplets}) 
in order to write the plaquette action concisely. 
Now we will rather decompose the full action into a sum over \emph{links},
which forces us to gather the auxiliary bosonic fields 
into two series of \emph{site-link multiplets}, in order to include
all fields coupled by $U\npmuh$ or $U^\dagger\npmuh$. 
We thus define two chirally-conjugated multiplets
associated with the site $n$ and the link $(n + \hmu/2)$:

\begin{equation}        \label{link-multiplets}
\Psi(n;\mu)  \defi  \begin{pmatrix} 
{\phi}^{(\mu,\mu+1)}_{R}(n)\\ 
{\phi}^{(\mu,-\mu-1)}_{L}(n)\\
\dots\\
{\phi}^{(\mu,d)}_{R}(n)\\
{\phi}^{(\mu,-d)}_{L}(n)\\
{\phi}^{(\mu,-1)}_{R}(n)\\
{\phi}^{(\mu,1)}_{L}(n)\\
\dots\\
{\phi}^{(\mu,-\mu+1)}_{R}(n)\\ 
{\phi}^{(\mu, \mu-1)}_{L}(n)
\end{pmatrix};\qquad 
\Phi(n;\mu)  \defi  \begin{pmatrix} 
{\phi}^{(\mu,\mu+1)}_{L}(n)\\ 
{\phi}^{(\mu,-\mu-1)}_{R}(n)\\
\dots\\
{\phi}^{(\mu,d)}_{L}(n)\\
{\phi}^{(\mu,-d)}_{R}(n)\\
{\phi}^{(\mu,-1)}_{L}(n)\\
{\phi}^{(\mu,1)}_{R}(n)\\
\dots\\
{\phi}^{(\mu,-\mu+1)}_{L}(n)\\ 
{\phi}^{(\mu, \mu-1)}_{R}(n)
\end{pmatrix}
\end{equation} 
The multiplets $\Psi(n;-\mu)$ and $\Phi(n;-\mu)$ 
associated with the site $n$ and link $n-\hmu/2$ are obtained from the ones
above by flipping the \emph{first} superscript $\mu$ into $-\mu$ in all
components, while changing neither the second superscript nor the ordering of 
the fields.

\begin{figure}[t]
\begin{center}
  \setlength{\unitlength}{1cm}
\begin{picture}(13,7.5)
  \put(1.5,0.5) {\mbox{\epsfysize=8.0cm\epsffile{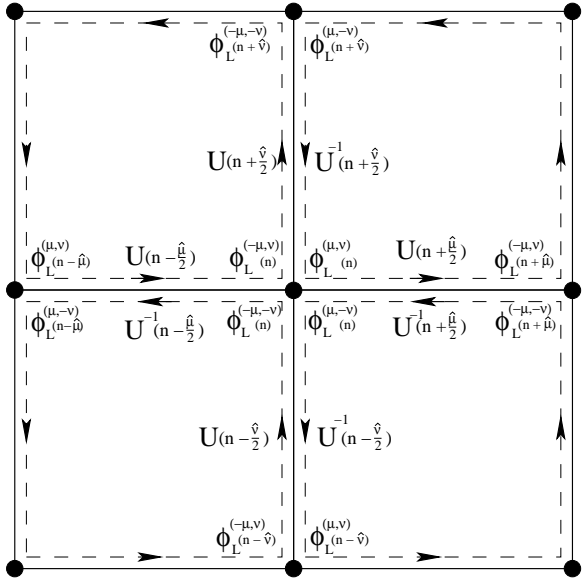}}}
  \lbfig{ind-link}
\end{picture}
\caption{Gauge couplings of the ``left'' bosonic fields carried by the
$4$ plaquettes in the plane  $(\mu\nu)$ around the lattice site $n$.
}
\end{center}
\end{figure}

Recall that the fields
$\phi^{(\mu,\nu)}$ are vectors with respect to the color group.   
Thus, the link multiplets 
$\Phi_a^i (n; \mu),\ \Psi_a^i (n;\mu)$ have to be labeled by      
color ($i$) and flavor ($a$) indices. The latter are given by the
second superscript in the definition of the multiplets: for instance, the
flavor indices of the multiplets \eqref{link-multiplets}
take the successive values $a = \mu+1,\ -\mu-1$, $\mu+2,\ldots$ etc.
Had we included multiple generations, the dimension of the 
coupling matrices would have been $d'=n_b(2d-2)$). 

Since there are $2d$ links around the site $n$ and for each link, two 
multiplets containing $2d-2$ components, the full set of multiplets is
of dimension $8d(d-1)$. On the other hand, the number of independent 
flavor components at each site is $2\times N_b=4d(d-1)$ (the factor $2$ 
corresponds to the chirality). Therefore, the site-link multiplets are not
linearly independent; indeed, each field component 
$\phi^{(\pm\mu,\pm\nu)}_{L/R}(n)$
is contained in exactly two multiplets, one associated with the link 
$(n+\hmu/2)$, the other with the link $(n+\hnu/2)$.

In terms of these multiplets, the interacting part of the action 
\eqref{z-boson} associated with the link $(n+\hmu/2)$
can be written in a compact form as follows (repeated indices are
summed over):
\begin{equation} \label{link-action-mult}
-S_U\npmuh =  {\bar\Phi}_{a}^{i}(n + \hmu ; -\mu) U^{ij}\npmuh 
\Phi^{j}_{a}(n;\mu) + 
{\bar\Psi}_{a}^{i}(n;\mu)  U^{\dagger ij}\npmuh \Psi^j_a(n+ \hmu; -\mu). 
\end{equation}
 
\medskip

Now that we isolated the part of the action associated with the matrix
$U\npmuh$, we can apply the bosonic $U(N_c)$ color-flavor transformation 
on this action, that is
replace the integration over $U\npmuh$ by
an integral over a complex matrix  $Z\npmuh$ of dimension
$d'$ \cite{Zirn}:
\begin{multline}   
\label{cft-bar}
\int_{\Gr[N_c]{U}} dU\npmuh \exp\left[-S_U\npmuh\right]
= \int_{D_{d'}}\! d\mu(Z,  Z^\dagger)\;  {\det(1-Z Z^\dagger)^{N_c}}\times\\  
\times\exp\biggl[{\bar \Phi}_{a}^{i}(n+\hmu;-\mu)
    Z_{ab} \npmuh \Psi^i_b(n+\hmu; -\mu) + 
{\bar \Psi}_{b}^i (n;\mu) {Z^\dagger}_{ba}\npmuh \Phi_{a}^i(n;\mu)\biggr].
\end{multline}
$D_{d'}$ denotes the set of complex matrices $Z$ of dimension $d'$
such that the Hermitian matrix $1-ZZ^\dagger$ is positive definite. This
set is in one-to-one correspondence with the non-compact symmetric space 
$\Gr[d',d']{U}/\Gr[d']{U}\times\Gr[d']{U}$ \cite{Zirn}, and the measure
$d\mu(Z,Z^\dagger)$ is the (suitablly normalized) invariant measure 
on this symmetric space: 
$$
d\mu(Z,Z^\dagger)=const\times\det(1-ZZ^\dagger)^{-2d'}
\prod_{a,b=1}^{d'}dZ_{ab}\,d\bar Z_{ab}.
$$ 
The identity \eqref{cft-bar} makes sense iff
\begin{equation}
N_c\geq 2 d'=4(d-1),
\end{equation} 
otherwise the integral over $Z$ does not converge.

In the color-flavor transformed action, the auxiliary fields are
coupled \emph{ultralocally} via the $Z$ matrices through 
their flavor indices. The indices of the matrix $Z\npmuh$ 
are associated with the plaquettes adjacent to the link $(n+\hmu/2)$, 
so that each entry of that matrix describes a correlation
between these plaquettes. This is to be put in contrast with 
the original 
action \eqref{z-boson}, which described a parallel transport of the
bosonic field along the links. In other 
words, if the $U$-field is responsible for the
transport along the link, the Z-field is responsible for the correlations
between different plaquettes. We shall see that the latter correlations 
become suppressed in the large mass limit.

Grouping all auxiliary fields at the site $n$ we get the interaction
part of the local effective action  
\begin{equation}            \label{SZn}
-S_Z [n] = \sum_{\mu=1}^d \left[  
\bar\Psi^i_b(n;\mu) {\bar Z}_{ab} \npmuh \Phi^i_a(n;\mu)
+\bar\Phi^i_a(n;-\mu) Z_{ab}\nmmuh \Psi^i_b(n;-\mu)\right],
\end{equation}
which is diagonal in the color indices.
Since the mass term
is diagonal with respect to both the color and
flavor indices, the color degrees of freedom are decoupled
in the transformed action, so that the partition function can be
factorized into $N_c$ identical integrals, each corresponding to a given
color.
Still, the coupling between the
$4d(d-1)$ flavor components at each site is not completely obvious, so we 
first analyze the simpler case of $d=2$ before turning to the general case. 


\subsection{d=2 effective action}
The simplest possible situation corresponds to the model placed 
on the 2-dimensional square lattice 
spanned by two orthogonal unit vectors $\hat 1$ and $\hat 2$. Let us 
consider the four links having the lattice site $n$ in common 
(see Fig~\ref{ind-link}). 
The space of auxiliary fields at $n$ is of dimension $8$ and, 
according to (\ref{SZn},\ref{link-multiplets}), the site-link multiplets 
(here, doublets) read
\begin{equation} \label{link-d=2}
\begin{split}
\Psi(n;1)  &=  \binom{\phi^{(1,2)}_R(n)}{\phi^{(1,-2)}_L(n)};\qquad 
\Psi(n;-1) = \binom{\phi^{(-1,2)}_R(n)}{\phi^{(-1,-2)}_L(n)}\\
\Psi(n;2)  &=  \binom{\phi^{(2,-1)}_R(n)}{\phi^{(2,1)}_L(n)};\qquad 
\Psi(n;-2)  = \binom{\phi^{(-2,-1)}_R(n)}{\phi^{(-2,1)}_L(n)}.
\end{split}
\end{equation}
The $4$ chirally conjugated doublets $\Phi(n;\pm\hat\alpha)$ are obtained
by exchanging $L\leftrightarrow R$.
These doublets are coupled through the $2\times 2$ matrices
$Z^\dagger\npoh, Z^\dagger\npth, Z\nmoh$ and 
$Z\nmth$ carried by the four links adjacent to the site $n$. 
To give an example, the matrix $Z^\dagger\npoh$ has the following index 
structure:  
\begin{equation}   \label{Z-entry-2}
Z^\dagger \npoh = \begin{pmatrix} 
Z^\dagger_{2,2}\npoh& 
Z^\dagger_{2,-2}\npoh\\
Z^\dagger_{-2,2}\npoh&
Z^\dagger_{-2,-2}\npoh
\end{pmatrix}
\end{equation}
Together with the link carrying the matrix, the lower pair of
indices represent the plaquettes associated with the field components 
coupled by the matrix element: the 
diagonal element $Z^\dagger_{2,2}\npoh$ couples different fields 
associated with
the same plaquette $(n; 1,2)$, while the nondiagonal element
${Z^\dagger}_{-2,2}\npoh$ couples fields associated with the two plaquettes
$(n; 1,2)$ and $(n; 1,-2)$.

\medskip

We want to write an effective action uniquely in 
terms of the $Z$ fields, by integrating over the bosonic fields. 
For this aim, we need to describe the coupling between each pair 
or flavors in the action~\eqref{SZn}.
As we already mentioned, the   
site-link multiplets \eqref{link-d=2} are not independent of each
other, so we now group the 
auxiliary fields at the lattice site $n$ 
into chirally conjugated \emph{site quadruplets}: 
\begin{equation}         \label{2-site-quadruplets}
\Phi(n) \defi  \begin{pmatrix} 
{\phi}^{(1,2)}_{R}(n)\\ 
{\phi}^{(1,-2)}_{L}(n)\\
{\phi}^{(-1,2)}_{L}(n)\\
{\phi}^{(-1,-2)}_{R}(n)
\end{pmatrix};\qquad 
\Psi(n) \defi \begin{pmatrix} 
{\phi}^{(1,2)}_{L}(n)\\ 
{\phi}^{(1,-2)}_{R}(n)\\
{\phi}^{(-1,2)}_{R}(n)\\
{\phi}^{(-1,-2)}_{L}(n)  
\end{pmatrix}.
\end{equation}
The union of these two quadruplets 
contain each bosonic component once. The 
color-flavor transformed action (\ref{SZn}) can be written in terms
of these quadruplets 
via two complex $4\times 4$ matrices 
in the flavor space, $V(n)$ and $W(n)$, 
which contain the components of the $Z$-fields:
\begin{equation}   \label{S-site-d2}
-S_Z [n] = \Phi^\dagger(n) V(n) \Psi(n)
+ \Psi^\dagger(n) W(n) \Phi(n).
\end{equation}

The matrices $V(n)$ and $W(n)$ can be 
compactly written 
\begin{equation}   \label{VW-fields}
V(n) \defi \begin{pmatrix} Z^\dagger \npoh&0\\
0& Z\nmoh\end{pmatrix};
\qquad W(n) \defi 
\tau_{(1,4)} \begin{pmatrix}Z\nmth &0\\ 0& Z^\dagger \npth \end{pmatrix}
\tau_{(1,4)}
\end{equation}
where the permutation matrix $\tau_{(1,4)}$ interchanges the first and 
fourth indices. In other words, 
we make use of the basis elements of the simple 
matrix algebra of order 2  
\begin{equation}   \label{T-generators}
T_1 \equiv e_{11} = \begin{pmatrix}1&0\\0&0
\end{pmatrix} \quad T_2 \equiv e_{22} = \begin{pmatrix}0&0\\0&1
\end{pmatrix},\quad T_3 \equiv e_{21} = \begin{pmatrix}0&0\\1&0
\end{pmatrix},\quad T_4 \equiv e_{12} = \begin{pmatrix}0&1\\0&0
\end{pmatrix}
\end{equation}
which satisfy 
$$
e_{ij}e_{kl} = \delta_{jk} e_{il},\qquad i,i,k,l = 1,2
$$
Then the matrices  $V(n)$ and $W(n)$ 
can be expressed via the direct tensor products of first two of 
these generators and  
$Z$-fields for the `$\hat 1$' and `$\hat 2$' directed links respectively: 
\begin{equation}   \label{VW-fields-1}
V \defi T_1 \otimes Z^\dagger \npoh  + T_2 \otimes \nmoh;
\qquad W \defi 
 \tau_{(1,4)} [T_1 \otimes Z\nmth   + T_2\otimes Z^\dagger\npth ]\tau_{(1,4)}^{-1} 
\end{equation}
where the matrix 
\begin{equation}         \label{t2}
 \tau_{(1,4)}= \tau_{(1,4)}^{-1} = T_1 \otimes T_2 + T_2 \otimes T_1 + T_3 \otimes T_3
+ T_4 \otimes T_4 \equiv 
\begin{pmatrix}  0&0&0&1\\
0&1&0&0\\
0&0&1&0\\
1&0&0&0
\end{pmatrix}
\end{equation}
permutes the first and the fourth components of the site 
quadruplets (\ref{2-site-quadruplets}).

The integral over auxiliary fields at the site $n$ (including
one color component) reads
\begin{equation}       \label{Z=d2}
\begin{split}
\mathcal{Z}[n] =&\int d\Psi^\dagger(n)d\Psi(n)  
d\Phi^\dagger(n) d\Phi(n)\ \exp 
\biggl[-m_b (\Psi^\dagger\Psi + \Phi^\dagger\Phi)
+ \Phi^\dagger V \Psi + \Psi^\dagger W\Phi\biggr]\\
&\propto \Det\begin{pmatrix} m_b&-V\\-W&m_b\end{pmatrix}^{-1}
= \Det(m_b^2 - VM)^{-1}\\
&=\exp\left[-\Tr \ln (1 - m_b^{-2}V W)\right]  \approx 
\exp\left[m_b^{-2} \Tr\big(VW\big)\right].
\end{split}
\end{equation}
In the last line we expanded the logarithm to first order in $1/m_b$. 

The trace of the product $VW$ can be easily computed:
\begin{multline}\label{trace-d2}
\Tr\big(V(n)W(n)\big) = Z^\dagger_{2,2}\npoh Z^\dagger_{1,1}\npth
+ Z^\dagger_{-2,-2}\npoh Z_{1,1}\nmth\\ 
+ Z_{2,2}\nmoh Z^\dagger_{-1,-1}\npth +  Z_{-2,-2}\nmoh Z_{-1,-1}\nmth.
\end{multline}

\begin{figure}[t]
\begin{center}
  \setlength{\unitlength}{1cm}
\begin{picture}(13,7.0)
  \put(1.5,0.5) {\mbox{\epsfysize=8cm\epsffile{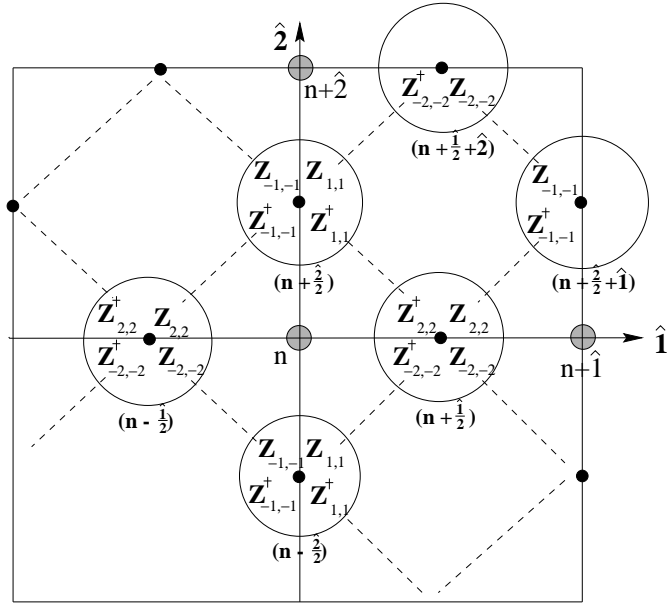}}}
  \lbfig{dual-2d}
\end{picture}
\caption{Schematic representation of the quadratic action \eqref{trace-d2}.
Each large circle contains diagonal matrix elements situated at some link.
The correlations between matrix elements of different links are depicted by
broken lines.}
\end{center}
\end{figure}

Notice that only the diagonal elements of the $Z$ matrices appear in
this leading-order term, which represent couplings between 
auxiliary fields carried by the same plaquette. In each term of the sum
\eqref{trace-d2}, the two matrix elements are 
carried by different links, but they correspond
to fields related to the same plaquette, precisely
the plaquette which shares these two links.
One can represent 
the correlations embodied in \eqref{trace-d2} 
by links of \emph{dual lattice} joining the
middles of the two coupled links of the original lattice (see Fig.~\ref{dual-2d}). 

Clearly, the two-dimensional model is self-dual because both the original and 
dual lattices are square ones. 

To summarize, to leading order 
in $1/m_b$ the full partition function is given by:
\begin{equation}\label{cft-part}
\mathcal{Z} =  
\int\left\{\prod_{n}\prod_{\alpha=1,2} \, d\mu(Z, Z^\dagger\npalh)\right\} \, 
\exp(-N_c\, S [Z]),
\end{equation}
with the effective action depending on the ``flavor'' matrices $Z$:
\begin{equation}       \label{cft-bar-z}
-S[Z]  = 
\sum_n  \biggl[ 
m_b^{-2} \Tr \bigl(V(n) W(n)\bigr) + \sum_{\alpha=1,2} 
\Tr\ln \left(1 -  Z\npalh Z^\dagger\npalh\right)\biggr] .
\end{equation}

\subsection{Saddle point equations and continuum limit of 
the $d=2$ effective theory}
The appearance of the factor $N_c$ in the exponent of the partition function
\eqref{cft-part} naturally suggests to imply the large-$N_c$ limit. 
Note that this limit is also needed to provide convergence of    
the bosonic color-flavor transformation, which is well-defined only if  
$N_c \ge 2N_b = 4 d(d-1)$. Thus, in the color-flavor 
transformed model the large-$N_c$ limit  becomes natural. Indeed,
the structure of the action  \eqref{cft-bar-z} suggests that in the 
large mass limit the functional integral over the  $Z,\, Z^\dagger$ 
fields is sharply 
peaked about the matrices close to unity, that is we can make use of   
the scalar ansatz $Z = z\mathbb{I};~~ Z^\dagger = z^*\mathbb{I}$, with $z$ and $z^*$ 
two scalar functions to evaluate the partition function in the saddle point 
approximation.

After inserting this ansatz into  \eqref{trace-d2}
we obtain 
\begin{equation}           \label{trace-d2-scalar}
\Tr\big(V(n)W(n)\big) =  2\, [z^*  \npoh +  z  \nmoh ]\, [z^* \npth + z \nmth] 
\end{equation}
and the action on the scalar ansatz becomes 
\begin{equation}   \label{act-z}
S[z] = 2 \sum\limits_n  \biggl[
\sum\limits_{\alpha=1,2} \ln  \left[1 -  
z{\ensuremath{({\scriptstyle n+\frac{\hat\alpha}{2}})}}z^* 
{\ensuremath{({\scriptstyle n+\frac{\hat\alpha}{2}})}}\right]
+ {m_b}^{-2} [z^*  \npoh +  z  \nmoh ]\, [z^* \npth + z \nmth] 
 \biggr] .
\end{equation}

In varying the action \eqref{act-z}, the
variables $z$ and $z^*$ are to be considered as
independent, which 
leads to two sets of saddle point equations 
\begin{equation}           \label{d2-saddle}
\begin{split}
z^* \npth + z\nmth &= \frac{m_b^2~ z\npoh}{1 - zz^*\npoh} = 
 \frac{m_b^2 ~z^*\nmoh}{1 - zz^*\nmoh};\nonumber\\
z^* \npoh + z\nmoh &= \frac{m_b^2~ z\npth}{1 - zz^*\npth} = 
 \frac{m_b^2 ~z^*\nmth}{1 - zz^*\nmth},
\end{split}
\end{equation}
which can be resolved if $z$ is just a spacetime--independent real number:
\begin{equation}           \label{z-saddle-d2}
z = z^* = \sqrt{1-\frac{m_b^2}{2}} 
\end{equation}
There is also a trivial solution $z = z^* = 0$.

Thus, the auxiliary scalar field cannot be set too heavy
since the effective action on the saddle point configuration 
$$
-S_{saddle}^{d=2} = 2 \left(\ln \frac{m_b^2}{2} - 1 + \frac{2}{m_b^2}\right) 
$$
becomes imaginary as $m_b^2 > 2$ in units of lattice spacing $a$. However,
even if we restrict the mass of inducing fields to the interval $ 1< m_b<2$,  
the corresponding induced coupling $\beta$ remains small enough 
to justify the large mass expansion. 

Let us consider a continuum limit of the model \eqref{cft-bar-z}.
We again use the scalar ansatz for the Z-fields to write the action in the 
form (\ref{act-z}). 

Following \cite{DP2000},  we can redefine the variables $z\npoh, z\npth$, 
which appears in the \emph{middle} of the links $\hat 1, \hat 2$, as differences
of $\omega(n)$ taken at the neighboring sites: 
\begin{equation} \label{omega-def}
\begin{split}
z\npoh &= \omega(n+1,2) - \omega(1,2) \approx 
a \partial_x \omega(n) + \frac{a^2}{2} \partial_x^2 \omega(n) + \dots;
\nonumber\\
z\nmoh &= \omega(1,2) - \omega(n-1,2) \approx 
a \partial_x \omega(n) - \frac{a^2}{2} \partial_x^2 \omega(n) + \dots;
\end{split}
\end{equation}
etc. Here the coordinate $x$ is taken along positive direction of the link $\hat 1$ 
and the coordinate $y$ is taken alon positive direction of the link  $\hat 2$. 

The Jacobian of this transformation is given by the determinant of 
$2\times 2$ matrix composed of the second  
derivatives: $J(\omega) = \det (\partial_x \partial_y \omega)$. 
In terms of these new variables one  
easily obtains 
\begin{equation} 
\begin{split}
\sum\limits_{\alpha = \hat 1,\hat 2} \ln 
 \left[1 -  z\npalh z^*\npalh\right]  
\approx \sum\limits_{\alpha = x,y} \ln(1 + a^2 \partial_\alpha 
\omega  \partial_\alpha \omega^*)   
& \approx
a^2 \sum\limits_{\alpha= x,y}(\partial_\alpha \omega) 
(\partial_\alpha \omega^*);
\nonumber\\
[z^*  \npoh +  z  \nmoh ]\, [z^* \npth + z \nmth] 
&\approx 2a^2 ~{\rm Re}~ \partial_x \omega \partial_y \omega . 
\end{split}
\end{equation}

Recall that in the  $d=2$ model the correlation length $\xi = m_b^{-1}$ remains finite 
in the continuum limit. Then the effective continuum coupling  $\beta = 
m_b^{-2}$ also is finite and 
in the continuum limit the effective two-dimensional action becomes 
\begin{equation}        \label{action-cont} 
\begin{split}
-S &= a^2 \sum\limits_{n}\big( \sum\limits_{\alpha=x,y}\,  
(\partial_\alpha \omega) (\partial_\alpha \omega^*)\big)
+  2\,  m_b^{-2}~{\rm Re}~ \partial_x \omega \partial_y \omega \\
&\to \int ~d^2x~ \left[ 
(\partial_x \omega)  
(\partial_x \omega^*) + (\partial_y \omega)  
(\partial_y \omega^*) + 
2\, \beta~{\rm Re}~ \partial_x \omega \partial_y \omega \right]
\end{split}
\end{equation}
where we restore conventional coordinate notations $\mu \to x$ and 
$\nu \to y$ in the continuum form of the effective action. 
Expression \eqref{action-cont} represents a dual version of d=2 
inducing theory in the large-$N_c$ strong coupling limit. 
Recall that the $Z$ field appears after the color-flavor transformation and 
therefore the action \eqref{action-cont} by definition 
does not contain color degrees of freedom. 
One may consider it as the action 
of the $\sigma$-model perturbed by the non-relativistic interaction term with a 
coupling constant $\beta$.  

Furthermore, we can use the method of Ref. \cite{DP2000} and identify the functions 
$\omega$ with the external coordinates of some manifold. The 
metric tensor  of the manifold is $g_{ij} = \partial_i\omega\, \partial_j\omega^*$ and 
the determinant of it is
$$
\det g_{ij} \equiv g =  \frac{1}{2}\varepsilon^{ij}\varepsilon^{kl} 
(\partial_i\omega\, \partial_k\omega^*)( \partial_j\omega\, \partial_l\omega^*).  
$$
Then the kinetic term of the continuum action \eqref{action-cont} can be 
represented in terms of the metric tensor as $(\partial_i \omega)  
(\partial_i \omega^*) = g_{ii}$ in correspondence with Ref. \cite{DP2000}. However,  
there is no straihgtforward generalization of the above considered continuum 
limit of $d=2$ model for a higher dimensional case.  

\subsection{$d=3$ effective action}
\begin{figure}[t]
\begin{center}
  \setlength{\unitlength}{1cm}
\begin{picture}(13,6.5)
  \put(-1.5,0.5) {\mbox{\epsfysize=5.8cm\epsffile{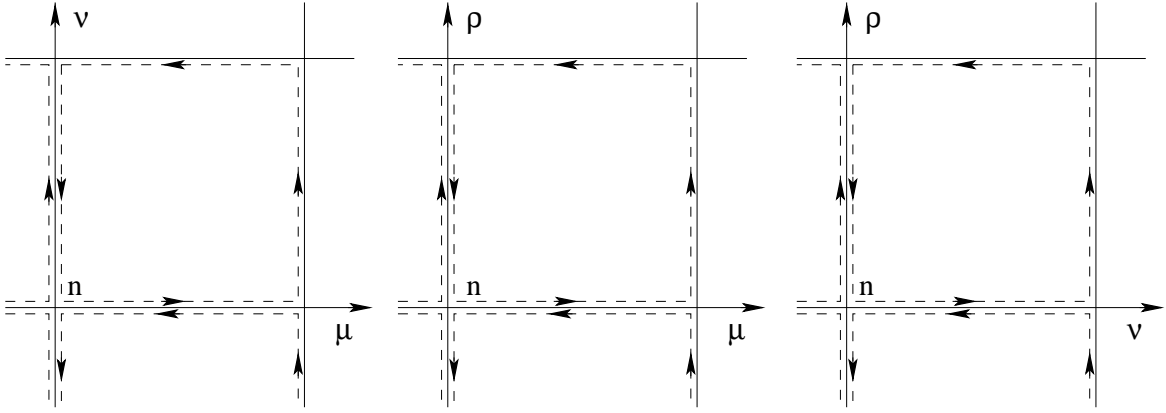}}}
  \lbfig{plaq-3d}
\end{picture}
\caption{Definition of the ``left'' chiral components on the planes 
of $d=3$ lattice.
}
\end{center}
\end{figure}
The consideration of the model on
3-dimensional lattice spanned by three vectors 
$\hmu,\hnu,{\hat \rho}$ becomes more complicated since the 
space of auxiliary fields is now of dimension of 24. 
First, we have to fix the
orientation of the plaquettes in 3 planes 
$(\mu\nu),(\mu\rho)$ and $(\nu\rho)$ according to the convention 
(\ref{left-convention}). We suppose that $\mu \mapsto 1$, 
$\nu \mapsto 2$ and  $\rho \mapsto 3$, thus the left chiral  components 
run on these plaquettes as in Fig. \ref{plaq-3d}. 

Let us analyze 
the structure of the action (\ref{SZn}) in that particular case. 
According to the 
definition of the link multiplets  (\ref{link-multiplets}), there 
are 6 scalar quadruplets whose explicit form is
\begin{equation}          \label{link-d=3}
\begin{split}
\Psi(n;1)  &\defi  \begin{pmatrix}
{\phi}^{(1,2)}_{R}(n)\\
{\phi}^{(1,-2)}_{L}(n)\\
{\phi}^{(1,3)}_{R}(n)\\
{\phi}^{(1,-3)}_{L}(n)
\end{pmatrix};\quad
\Psi(n;2)  \defi  \begin{pmatrix}
{\phi}^{(2,3)}_{R}(n)\\
{\phi}^{(2,-3)}_{L}(n)\\
{\phi}^{(2,-1)}_{R}(n)\\
{\phi}^{(2,1)}_{L}(n)
\end{pmatrix};\quad
\Psi(n;3)  \defi  \begin{pmatrix}
{\phi}^{(3,-1)}_{R}(n)\\
{\phi}^{(3,1)}_{L}(n)\\
{\phi}^{(3,-2)}_{R}(n)\\
{\phi}^{(3,2)}_{L}(n)
\end{pmatrix};\nonumber\\
\Phi(n;1)  &\defi  \begin{pmatrix}
{\phi}^{(1,2)}_{L}(n)\\
{\phi}^{(1,-2)}_{R}(n)\\
{\phi}^{(1,3)}_{L}(n)\\
{\phi}^{(1,-3)}_{R}(n)
\end{pmatrix};\quad
\Phi(n;2)  \defi  \begin{pmatrix}
{\phi}^{(2,3)}_{L}(n)\\
{\phi}^{(2,-3)}_{R}(n)\\
{\phi}^{(2,-1)}_{L}(n)\\
{\phi}^{(2,1)}_{R}(n)
\end{pmatrix};\quad
\Phi(n;3)  \defi  \begin{pmatrix}
{\phi}^{(3,-1)}_{L}(n)\\
{\phi}^{(3,1)}_{R}(n)\\
{\phi}^{(3,-2)}_{L}(n)\\
{\phi}^{(3,2)}_{R}(n)
\end{pmatrix},
\end{split}
\end{equation}

These multiplets are coupled with 6 
matrix $Z$-fields of dimension $4\times 4$
defined on the 6  directed links $\hat 1$, $\hat 2$ and $\hat 3$ which are 
adjacent to the point $n$: 
\begin{equation}            \label{SZn-d3}
-S_Z [\Psi(n), \Phi(n)]  =   \sum\limits_{\alpha = 1}^3 
\left[{\bar \Psi} (n;\alpha) {Z^\dagger} \npalh \Phi (n;\alpha)
+ {\bar \Phi} (n;-\alpha) Z\nmalh \Psi(n; -\alpha) \right]
\end{equation}
The explicit form of the matrix $Z^\dagger \npoh$ for example is  (cf. 
its $2d$ counterpart (\ref{Z-entry-2}))
\begin{equation}        \label{Z-entry-3}
Z^\dagger \npoh = \begin{pmatrix} 
 Z^\dagger_{2,2} \npoh & 
 Z^\dagger_{2,-2}\npoh &
 Z^\dagger_{2,3} \npoh &
 Z^\dagger_{2,-3}\npoh \\
 Z^\dagger_{-2,2}\npoh &  
 Z^\dagger_{-2,-2}\npoh &
 Z^\dagger_{-2,3} \npoh &
 Z^\dagger_{-2,-3}\npoh \\
 Z^\dagger_{3,2} \npoh &
 Z^\dagger_{3,-2}\npoh &
 Z^\dagger_{3,3} \npoh &
 Z^\dagger_{3,-3}\npoh \\
 Z^\dagger_{-3,2} \npoh &
 Z^\dagger_{-3,-2} \npoh &
 Z^\dagger_{-3,3} \npoh &
 Z^\dagger_{-3,-3}  \npoh &
\end{pmatrix}~,~~ {\rm etc}
\end{equation}

To simplify the evaluation of the functional determinant of the color-flavor
transformed model in $d=3$ we have to bring the matrix, 
which couples all these multiplets and contains different components
of the $Z$ fields, to the block-diagonal form.
Thus we again regroup the components of the scalar link multiplets
(\ref{link-d=3}) into the site quadruplets: 
\begin{equation}         \label{site-multiplets-d3}
\begin{split}
\Phi^{(1 2)} &\defi  \begin{pmatrix} 
{\phi}^{(1,2)}_{R}\\ 
{\phi}^{(1,-2)}_{L}\\
{\phi}^{(-1,2)}_{L}\\
{\phi}^{(-1,-2)}_{R}
\end{pmatrix};\qquad
\Phi^{(13)} \defi  \begin{pmatrix} 
{\phi}^{(1,3)}_{R}\\ 
{\phi}^{(1,-3)}_{L}\\
{\phi}^{(-1,3)}_{L}\\
{\phi}^{(-1,-3)}_{R}
\end{pmatrix};\qquad
\Phi^{(23)} \defi  \begin{pmatrix} 
{\phi}^{(2,3)}_{R}\\ 
{\phi}^{(2,-3)}_{L}\\
{\phi}^{(-2,3)}_{L}\\
{\phi}^{(-2,-3)}_{R}
\end{pmatrix} \\
{\Psi}^{(12)} &\defi \begin{pmatrix} 
{\phi}^{(1,2)}_{L}\\ 
{\phi}^{(1,-2)}_{R}\\
{\phi}^{(-1,2)}_{R}\\
{\phi}^{(-1,-2)}_{L}
\end{pmatrix};\qquad
{\Psi}^{(13)} \defi \begin{pmatrix} 
{\phi}^{(1,3)}_{L}\\ 
{\phi}^{(1,-3)}_{R}\\
{\phi}^{(-1,3)}_{R}\\
{\phi}^{(-1,-3)}_{L}
\end{pmatrix};\qquad
{\Psi}^{(23)} \defi \begin{pmatrix} 
{\phi}^{(2,3)}_{L}\\ 
{\phi}^{(2,-3)}_{R}\\
{\phi}^{(-2,3)}_{R}\\
{\phi}^{(-2,-3)}_{L}
\end{pmatrix}
\end{split}
\end{equation}

In terms of these variables the local actions (\ref{SZn-d3}) 
can be rewritten as (cf. Eq.~(\ref{S-site-d2}))
\begin{equation}   \label{S-site-d3}
\begin{split}
-S_Z [n] &=  {\bar \Phi}^{(12)} V_{(1)}^{(2,2)} 
{\Psi}^{(12)} + {\bar{\Phi}}^{(13)} V_{(1)}^{(3,3)}
\Psi^{(13)} +  {\bar \Phi}^{(12)} V_{(1)}^{(2,3)} 
{\Psi}^{(13)}\\
&+ {\bar \Phi}^{(13)} V_{(1)}^{(3,2)} 
{\Psi}^{(12)} +  {\bar \Phi}^{(23)} V_{(2)}^{(3,3)} 
{\Psi}^{(23)} +  {\bar \Psi}^{(12)} W_{(2)}^{(1,1)} 
{\Phi}^{(12)} \nonumber\\
&+ {\bar \Psi}^{(13)} W_{(3)}^{(1,1)} 
{\Phi}^{(13)} + {\bar \Psi}^{(13)} W_{(3)}^{(1,2)} 
{\Phi}^{(23)} + {\bar \Psi}^{(23)} W_{(3)}^{(2,1)} 
{\Phi}^{(13)}\nonumber\\
&+ {\bar \Psi}^{(23)} W_{(3)}^{(2,2)} 
{\Phi}^{(23)} + {\bar \Phi}^{(23)} R_{(2)}^{(3,1)} 
{\Phi}^{(12)} + {\bar \Psi}^{(12)} S_{(2)}^{(1,3)} 
{\Psi}^{(23)}
\end{split}
\end{equation}
where the components of the $Z$ fields are now regrouped into new $4\times 4$ 
matrices which are coupled with the site quadruplets (\ref{site-multiplets-d3})
in the action (\ref{S-site-d3}). The lower index of these matrix fields is associated
with the link which carries the components of the $Z$ field,
while the upper pair represents the plaquette indices of the site 
quadruplets which cap the link. 

The matrices $V_{(\alpha)}^{(\beta,\gamma)},
~\alpha = 1,2;~ \beta,\gamma = 2,3$ are defined on the first two 
links and couple the multiplets $ {\bar \Phi}$ and $\Psi$; one of these 
matrices is the above defined d=2 couplig matrix $V(n) \equiv V_{(1)}^{(2,2)}$
which appears in \eqref{Z=d2}. The matrices $W_{(\alpha)}^{(\beta,\gamma)},
~\alpha = 2,3;~ \beta,\gamma = 1,2$ are defined on the second and 
third links and couple the multiplets $ {\bar \Psi}$ and $\Phi$, one 
of these matrices also appears in \eqref{Z=d2}. 
Unlike the $d=2$ model there are two more matrices  
$R_{(2)}^{(3,1)}$ and $S_{(2)}^{(1,3)}$ which live on the 
second links and couple the fields  ${\bar \Phi}, \Phi$ and  
${\bar \Psi}, \Psi$, respectively.  

These matrices can be written concisely by
making use of the above defined generators $T_{i}$ (\ref{T-generators}). 
Let us  supplement the permutation matrix $\tau_2$ of eq. (\ref{t2}) by 
two other matrices 
\begin{equation}         \label{tau}
\begin{split}
\tau_1 &= T_3  \otimes \mathbb{I} + 
+ T_4 \otimes \mathbb{I} \equiv 
\begin{pmatrix}  0&0&1&0\\
0&0&0&1\\
1&0&0&0\\
0&1&0&0
\end{pmatrix};\nonumber\\
\tau_3 &= T_1  \otimes T_4 + 
T_4  \otimes T_2 + T_2  \otimes T_3 + T_3  \otimes T_1 
\equiv 
\begin{pmatrix}  0&1&0&0\\
0&0&0&1\\
1&0&0&0\\
0&0&1&0
\end{pmatrix},
\end{split}
\end{equation}
which permute all four indices of the site quadruplets 
simultaneously, $(1,2,3,4) 
\rightleftharpoons 
(3,4,1,2)$ and $(1,2,3,4) \rightleftharpoons (3,1,4,2)$ respectively.

One can consider the basis elements of the matrix algebra (\ref{T-generators})
as generators whose tensor products with a  complex $2\times 2$ matrix define 
different embeddings into a matrix of dimension $4\times 4$. Then
each of the 6 matrices of the $Z$ field can be written 
as an expansion in the basis of the generators $T_i$:
$$
Z = \sum\limits_{i=1}^4 Z_i \otimes T_i 
$$
where $2\times 2$ blocks in components are given by 
\begin{equation}
Z_1 = \begin{pmatrix}Z_{11}&Z_{12}\\
                     Z_{21}&Z_{22}\end{pmatrix};\qquad
Z_2 = \begin{pmatrix}Z_{33}&Z_{34}\\
                     Z_{43}&Z_{44}\end{pmatrix};\qquad
Z_3 = \begin{pmatrix}Z_{31}&Z_{32}\\
                     Z_{41}&Z_{42}\end{pmatrix};\qquad
Z_4 = \begin{pmatrix}Z_{13}&Z_{14}\\
                     Z_{23}&Z_{24}\end{pmatrix}
\end{equation}

With all these shortnotes at hand, we can write
\begin{equation}   \label{V-and-W}
\begin{split}
V_{(1)}^{(2,2)} &= Z_1^\dagger  \npoh
+ \tau_1 Z_1\nmoh \tau_1^{-1}; \qquad 
V_{(1)}^{(3,3)} = \tau_1 Z_2^\dagger\npoh \tau_1^{-1}
+ Z_2 \nmoh;\nonumber\\
V_{(1)}^{(2,3)} &=  
Z_4^\dagger\npoh \tau_1^{-1} + \tau_1 Z_4 \nmoh; \qquad 
V_{(1)}^{(3,2)} =   
\tau_1 Z_3^\dagger\npoh + Z_3\nmoh\tau_1^{-1};\nonumber\\
V_{(2)}^{(3,3)} &= Z_1^\dagger\npth + 
\tau_1 Z_1\nmth \tau_1^{-1}; \qquad   
W_{(2)}^{(1,1)} = \tau_2 Z_2^\dagger\npth \tau_2^{-1} 
+ \tau_3 Z_2 \nmth \tau_3^{-1};\nonumber\\
W_{(3)}^{(1,1)} &=  \tau_3 Z_1^\dagger\npthh  \tau_3^{-1} 
+ \tau_2 Z_1\nmthh  \tau_2^{-1}; \qquad 
W_{(3)}^{(2,2)} = \tau_2  Z_2^\dagger\npthh \tau_2^{-1} 
+ \tau_3  Z_2 \nmthh  \tau_3^{-1};\nonumber\\
W_{(3)}^{(1,2)} & =  \tau_3  Z_4^\dagger\npthh \tau_2^{-1}
+ \tau_2 Z_4 \nmthh  \tau_3^{-1};\qquad 
W_{(3)}^{(2,1)} = \tau_2  Z_3^\dagger\npthh \tau_3^{-1}
+ \tau_3 Z_3 \nmthh \tau_2^{-1};\nonumber\\
 S_{(2)}^{(1,3)} &= \tau_2 Z_3^\dagger\npth + \tau_3   
Z_3\nmth  \tau_1^{-1};\qquad
  R_{(2)}^{(3,1)} = Z_4^\dagger\npth\tau_2^{-1} + 
\tau_1  Z_4\nmth \tau_3^{-1}  
\end{split}
\end{equation}

To complete our calculations and carry out the integration over 
the scalar fields, we compose the site multiplets  
(\ref{site-multiplets-d3}) into 12-plets    
$$
\Phi \defi  \begin{pmatrix}\Phi^{(12)}\\ 
\Phi^{(13)}\\
\Phi^{(23)}
\end{pmatrix};\qquad
\Psi \defi  \begin{pmatrix}\Psi^{(12)}\\ 
\Psi^{(13)}\\
\Psi^{(23)}
\end{pmatrix}
$$
which become coupled with the total coupling matrices. These matrices 
have the following block structure
\begin{equation}
V \defi \begin{pmatrix} V_{(1)}^{(2,2)}& V_{(1)}^{(3,2)}&0\\
                    V_{(1)}^{(2,3)}& V_{(1)}^{(3,3)}&0\\
0&0&V_{(2)}^{(3,3)}
\end{pmatrix};\qquad W \defi \begin{pmatrix} W_{(2)}^{(1,1)}&0&0\\
                         0&W_{(3)}^{(1,1)}&W_{(3)}^{(1,2)}\\
                      0&W_{(3)}^{(2,1)}&W_{(3)}^{(2,2)}
\end{pmatrix}.
\end{equation}

Clearly,  the mass term of the auxiliary fields 
needs to be taken into account in the final expression for the action.
Then the matrices $S_{(2)}^{(1,3)}$ and $R_{(2)}^{(3,1)}$
appear as off-diagonal blocks of the total mass matrices 
\begin{equation}
M_1 \defi \begin{pmatrix} m_b&0&0\\
			0&m_b&0\\
		       R_{(2)}^{(3,1)}&0&m_b
\end{pmatrix};\qquad M_2 \defi \begin{pmatrix} 
m_b&0&S_{(2)}^{(1,3)}\\
			0&m_b&0\\
		       0&0&m_b
\end{pmatrix}
\end{equation}
respectively. Note that $\Det M_1 = \Det M_2 = m_b^3$. 
  
Thus, the complete action can be written in the form 
$$
-S[n] = {\bar \Phi} V \Psi + {\bar \Psi} W \Phi + {\bar \Phi} M_1 
\Phi + {\bar \Psi} M_2 \Psi 
$$
and the functional integration over the auxiliary fields yields
the determinant of the full coupling matrix of dimension 24: 
\begin{equation}
\Det^{-N_c} \begin{pmatrix} M_1&V\\
                            W&M_2\end{pmatrix} =
\Det^{-N_c}(M_1M_2 - VW) \propto \Det^{-N_c}(1 - M_2^{-1}M_1^{-1}VW)
\end{equation} 

Since 
$$
M_1^{-1} = \begin{pmatrix} 
m_b^{-1}&0&0\\
			0&m_b^{-1}&0\\
		       -R_{(2)}^{(3,1)}m_b^{-2} &0&m_b^{-1}
\end{pmatrix};\qquad
M_2^{-1} = \begin{pmatrix} 
m_b^{-1}&0&-S_{(2)}^{(1,3)}m_b^{-2}\\
			0&m_b^{-1}&0\\
		       0&0&m_b^{-1}
\end{pmatrix} ,
$$
the leading order of the large mass expansion yields 
\begin{equation}
\Det^{-N_c}(1 - M_2^{-1}M_1^{-1}VW) \propto \exp\left[-N_c \Tr \ln (1-
\frac{1}{m_b^2}VW\right] \approx \exp\left[\frac{N_c}{m_b^2}\Tr(VW)\right] 
\end{equation}
where the explicit form of the trace is rather cumbersome
\begin{equation}           \label{trace-3d}
\begin{split}
\Tr\big(V(n)W(n)\big) &= 
Z^\dagger_{2,2} \npoh Z^\dagger_{1,1} \npth + 
Z_{2,2}\nmoh Z^\dagger_{-1,-1}\npth \\ 
&+ Z^\dagger_{-2,-2}  Z_{1,1}\nmth 
+ Z_{-2,-2}\nmoh   Z_{-1,-1}\nmth \\
&+ Z^\dagger_{1,1} \npthh Z^\dagger_{3,3}\npoh + 
 Z^\dagger_{-2,-2} \npthh  Z_{3,3} \nmoh \\
&+ Z^\dagger_{-3,-3} \npoh Z_{1,1}\nmthh + 
Z_{-1,-1}\nmthh  Z_{-3,-3}\nmoh \\
&+ Z^\dagger_{2,2} \npthh  Z^\dagger_{3,3}\npth 
+ Z^\dagger_{-3,-3}\npth Z_{2,2}\nmthh \\
&+ Z^\dagger{-2,-2}\npthh  Z_{3,3}\nmth 
+ Z_{-2,-2} \nmthh  Z_{-3,-3} \nmth
\end{split}
\end{equation} 

Note, that as in the $d=2$ model (cf. Eq.~(\ref{trace-d2}), 
in the leading order of the expansion in $1/m_b$  
only diagonal elements of the $Z$-matrix fields contribute to the 
functional determinant and 
the correlations between the plaquettes, 
which are described by \eqref{trace-3d} are factorized into the sum of 12  
independent two-points correlations between the diagonal components of the  
$Z$ matrix field. Therefore, the cube of the original d=3 lattice on the 
dual lattice corresponds to the \emph{tetradecahedra}   
whose edges corresponds to these correlations  
(see Fig. \ref{dual-3d}). The plaquette of the original lattice corresponds 
to a \emph{tetrahedron}, six of those precisely fit the tetradecahedron of a dual 
cube. This is exactly the structure suggested in \cite{DP2000} from a different viewpoint. 
However, the higher order corrections in $1/m_b^2$ make the geometry of the 
dual lattice much more complicated.
\begin{figure}[t]
\begin{center}
  \setlength{\unitlength}{1cm}
\begin{picture}(13,7.5)
  \put(1.5,0.5) {\mbox{\epsfysize=7.4cm\epsffile{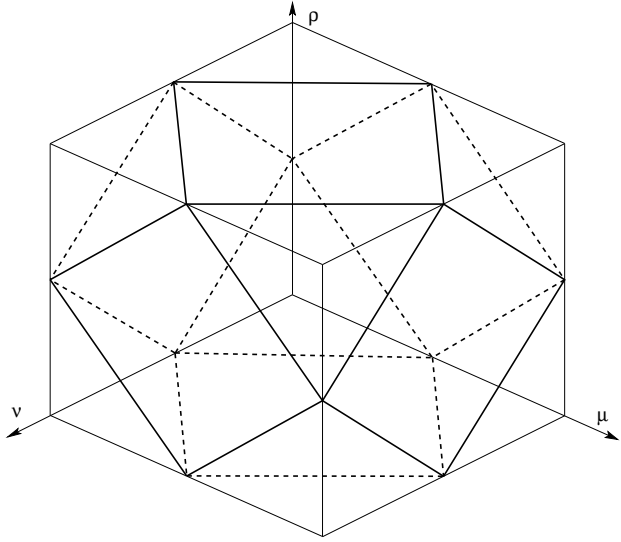}}}
  \lbfig{dual-3d}
\end{picture}
\caption{A tetradecahedron of the dual lattice 
corresponding to the cube of d=3 original lattice.  
}
\end{center}
\end{figure}

Finally, we arrive at the effective action of the 3d dual gluodynamics 
which has the same structure as \eqref{cft-bar-z}:
\begin{equation}       \label{cft-bar-z-d3}
-S[Z]  = 
\sum_n  \Tr \left[ 
m_b^{-2} V(n) W(n) + \sum\limits_{\alpha = 1}^3 
\ln 
 \left(1 -  Z\npalh Z^\dagger\npalh\right)\right] .
\end{equation}
We will see that this results holds in any dimension. 

Note that there is a relation between the space of the 
pairs of plaquette indices of $d$-dimensional space 
$(\alpha,\beta)$ where $\alpha,\beta = 1,2\dots d$, and the Cartan--Weyl \emph{simple root spaces}  
of two Lie  $\gr[d]{su}$-algebras which are subalgebras of $\gr[d]{gl}$.
Indeed, one can set a correspondence between 
the ordered pairs $(\alpha,\beta)$ with positive $\alpha$ and $\beta$, 
and the basis elements of the simple 
matrix algebra of $d$th order: 
\begin{equation}
\begin{split}
(\alpha,\beta) &\mapsto e_{\alpha\beta} ~~{\rm if}~~\alpha < \beta, \alpha,\beta > 0;
\nonumber\\
(-\alpha,-\beta) &\mapsto e_{\beta\alpha} ~~{\rm if}~~\alpha < \beta, \alpha,\beta > 0.
\end{split}
\end{equation}
Here the $d\times d$-matrices $e_{\alpha\beta}$ are d-dimensional generalization 
of the two-dimensional basis \eqref{T-generators}: the entry in the $\alpha$th row 
and $\beta$th column of $e_{\alpha\beta}$ is equal to 1 while all other entries are 
zero. With that ordering 
the matrices $e_{\alpha\beta}$ and $e_{\beta\alpha}$ can be identified as the  
raising and lowering generators  $E_{\pm \vec \beta_i}$ of an   $\gr[d]{su}$-algebra 
which is characterized by the set of the  
simple roots $\pm \vec \beta_i$ respectively. To cover all the space of the pairs of the 
plaquette indices we have to consider also the second set of the 
ordered pairs $(-\alpha,\beta)$ and 
$(\alpha,-\beta)$, which can be set into correspondence with a complimentary 
$\gr[d]{{\bar{su}}}$-algebra in the same way. 

For example, in the d=3 model with the above ordering 
of the indices,
the plaquettes $(n; 1,2)$ and  $(n; -1,-2)$ correspond to the matrices 
$$
(n; 1,2)\mapsto e_{12} = \begin{pmatrix} 0&1&0\\
                0&0&0\\
                0&0&0 \end{pmatrix} 
\equiv E_{\vec \beta_1};\quad (n; -1,-2)\mapsto e_{21} = \begin{pmatrix} 0&0&0\\
                1&0&0\\
                0&0&0 \end{pmatrix} 
\equiv E_{-{\vec \beta}_1}\, ,
$$
where $\vec \beta_1$ is one of the simple roots of the $\gr[3]{su}$-algebra. The second 
positive root $\vec \beta_2$ can be set into 
correspondence with the plaquette $(n; 2,3)$ whereas  
the third positive composite root $\vec \beta_3 = \vec \beta_1 + \vec \beta_2$ 
corresponds to the plaquette $(n; 1,3)$. 

To identify the raising and lowering generators of the second $\gr[3]{\bar{su}}$-algebra 
we have to consider, for example, the pairs of the plaquettes 
 $(n; 1, -2)$ and  $(n; -1,2)$ which can be mapped into the matrices ${\bar e}_{12}$ 
and ${\bar e}_{21}$ in the same way. This corresponds to the first simple root of the 
$\gr[3]{\bar{su}}$-algebra.   

This construction can be used to integrate over the 
auxiliary fields in a general d-dimensional case \cite{ShNonn}.

\subsection{Effective action in arbitrary dimension}
Our aim in this section is the same as in the previous two, that is 
integrate the action \eqref{SZn} over the auxiliary fields at the site $n$, 
and compute the resulting effective action in the matrices $Z$, $Z^\dagger$
in arbitrary dimension. We will only consider the case of
one generation of auxiliary fields, that is, $n_b=1$ \cite{ShNonn}.
As we already pointed out, the difficulty comes from the fact that the
same field $\phi^{(\pm\mu,\pm\nu)}_{L/R}$ is contained in two different
site-links multiplets \eqref{link-multiplets}. In order to 
integrate over these fields, we first need to regroup them, that is write
the action $S_Z$ as 
\begin{equation}
-S_Z[n]=\big[\phi^{(\pm\mu,\pm\nu)}_{L/R}\big]^\dagger 
\mathcal{M}(n)\big[\phi^{(\pm\mu,\pm\nu)}_{L/R}\big],
\end{equation}
where the column vector $\big[\phi^{(\pm\mu,\pm\nu)}_{L/R}\big]$ 
contains the $2N_b=4d(d-1)$
fields at the site $n$. The coupling matrix $\mathcal{M}(n)$ is therefore
of dimension $4d(d-1)$; it contains components of the matrices 
$Z$ and $Z^\dagger$
carried by the links touching $n$. Our task is to write down the matrix
$\mathcal{M}$ explicitly, using a judicious grouping of the field components. 
As was already the case
in two dimensions, the matrix $\mathcal{M}$ has many null entries, so that its
determinant may be simplified.

We will group the auxiliary fields in \emph{site quadruplets} associated with 
the planes in the $d$-dimensional lattice. Each plane, indexed by a 
couple $(\mu\nu)$ with
$1\leq\mu<\nu\leq d$, contains
$4$ plaquettes touching $n$. To each plane we associate two site quadruplets
at $n$:
\begin{equation}         \label{d-site-quadruplets}
\Phi^{(\mu\nu)}(n) \defi  \begin{pmatrix} 
{\phi}^{(\mu,\nu)}_{R}(n)\\ 
{\phi}^{(\mu,-\nu)}_{L}(n)\\
{\phi}^{(-\mu,\nu)}_{L}(n)\\
{\phi}^{(-\mu,-\nu)}_{R}(n)
\end{pmatrix};\qquad 
{\Psi}^{(\mu\nu)}(n) \defi \begin{pmatrix} 
{\phi}^{(\mu,\nu)}_{L}(n)\\ 
{\phi}^{(\mu,-\nu)}_{R}(n)\\
{\phi}^{(-\mu,\nu)}_{R}(n)\\
{\phi}^{(-\mu,-\nu)}_{L}(n)  
\end{pmatrix}.
\end{equation}
These quadruplets generalize the ones defined in 
Eq.~\eqref{2-site-quadruplets} to any plane in the $d$-dimensional lattice.
The total number of these planes is 
$\frac{d(d-1)}{2}$, so that the `concatenation' of all the above site 
quadruplets yields the correct number of field components. 
To perform this concatenation, we need to \emph{order} the different planes,
that is, to order the couples $(\mu\nu)$.

These couples 
are in one-to-one correspondence with the positive roots
of the Lie algebra $\gr[d]{gl}$: each plane $(\mu\nu)$ can indeed be 
associated with the generator $e_{\mu\nu}$ of the algebra, 
satisfying the relations
\begin{equation}\label{algebra}
[e_{\mu\nu},e_{\rho\eta}]
=\delta_{\nu\rho}e_{\mu\eta}-\delta_{\mu\eta}e_{\rho\nu}.
\end{equation}
There is no canonical ordering of the positive generators 
(or the positive roots), on the other hand it seems natural to require
that $(\mu\nu)< (\rho\eta)$ if $\nu\leq\rho$; this condition 
is satisfied by the following ordering:
\begin{equation}\label{convention}
(12)<(13)<(14)<\ldots<(1d)<(23)<(24)<\ldots<(2d)<(34)<\ldots<(d-1\,d).
\end{equation}
The site quadruplets will be ordered according to the above convention, 
starting with all quadruplets $\Phi^{(\mu\nu)}$ and finishing with the
quadruplets $\Psi^{(\mu\nu)}$.

Now that we ordered the vector $\big[\phi^{(\pm\mu,\pm\nu)}_{L/R}\big]$, 
we need to compute the matrix $\mathcal{M}$, and for this to
derive which quadruplets $\Phi^{(\mu\nu)\dagger}$ or $\Psi^{(\mu\nu)\dagger}$
are coupled with which quadruplets $\Phi^{(\rho\eta)}$ or 
$\Psi^{(\rho\eta)\dagger}$, and through which matrices $Z$ or $Z^\dagger$.
For this aim, we have to compare the components of, on one side, the
quadruplets $\Phi^{(\mu\nu)}$, $\Psi^{(\mu\nu)}$; on the other side, the 
site-link multiplets $\Phi(n;\pm \alpha)$, $\Psi(n;\pm \alpha)$ which were used
to write the action \eqref{SZn}. For instance, the first component 
$\phi^{(\mu,\nu)}_{R}$ of $\Phi^{(\mu\nu)}$ is contained in the site-link
multiplet $\Psi(n;\mu)$ (because $\mu<\nu$), 
so its complex conjugate is coupled to the matrix 
$Z^\dagger\npmuh$ on the left; on the other hand, 
$\phi^{(\mu,\nu)}_{R}$ is also
contained in the multiplet $\Phi(n;\nu)$ (because $\nu>\mu$), so it is
coupled to the matrix $Z^\dagger\npnuh$ on the right. Below we schematically 
represent the couplings of the quadruplets with the $Z$ matrices by taking all
components in the quadruplets into account:
\begin{equation}
\begin{split}
\Phi^{(\mu\nu)\dagger}\longrightarrow Z^\dagger\npmuh,\,Z\nmmuh&\qquad
Z^\dagger\npetah,\,Z\nmetah\longleftarrow\Phi^{(\rho\eta)}\\
\Psi^{(\mu\nu)\dagger}\longrightarrow Z^\dagger\npnuh,\,Z\nmnuh&\qquad
Z^\dagger\nprhoh,\,Z\nmrhoh\longleftarrow\Psi^{(\rho\eta)}.
\end{split}
\end{equation}
In general, one of the two indices in the couple $(\mu\nu)$ specifies the
direction of the link carrying the matrix $Z$ or $Z^\dagger$, while the other
index shows which entries of the matrix are concerned: for instance, 
$\bar\phi^{(\mu,\nu)}_R$ couples to the entries $Z^\dagger\npmuh_{\nu,.}$, 
while $\phi^{(\rho\eta)}_R$ couples to the entries $Z\nmetah_{.,\rho}$.

As a result, the couplings between the site quadruplets satisfy some 
``selection rules'', which mean that the matrix $\mathcal{M}$ contains many 
$4\times 4$ empty blocks. The non-empty blocks connect the following pairs of
quadruplets:
\begin{equation}\label{selection}
\begin{split}
\Phi^{(\mu\nu)\dagger}\longleftrightarrow\Phi^{(\rho\eta)}&\mbox{ iff }
\mu=\eta\\
\Phi^{(\mu\nu)\dagger}\longleftrightarrow\Psi^{(\rho\eta)}&\mbox{ iff }
\mu=\rho\\
\Psi^{(\mu\nu)\dagger}\longleftrightarrow\Phi^{(\rho\eta)}&\mbox{ iff }
\nu=\eta\\
\Psi^{(\mu\nu)\dagger}\longleftrightarrow\Psi^{(\rho\eta)}&\mbox{ iff }
\nu=\rho.
\end{split}
\end{equation}
By analogy with the $2$-dimensional case, we will call $V_\mu^{\nu,\eta}$
the matrix coupling $\Phi^{(\mu\nu)\dagger}$ with $\Psi^{(\mu\eta)}$, and
$W_{\nu}^{\mu,\rho}$ the matrix coupling $\Psi^{(\mu\nu)\dagger}$ with
$\Phi^{(\rho\nu)}$. The structure of these matrices is similar to the ones
in Eq.~\eqref{VW-fields}, except that the matrices $Z$, $Z^\dagger$ are 
replaced by $2\times 2$ submatrices:
\begin{equation}\label{VW-d}
\begin{split}
V_\mu^{\nu,\eta}(n)&=\begin{pmatrix}
Z^\dagger_{\nu,\eta}\npmuh&Z^\dagger_{\nu,-\eta}\npmuh&0&0\\
Z^\dagger_{-\nu,\eta}\npmuh&Z^\dagger_{-\nu,-\eta}\npmuh&0&0\\
0&0&Z_{\nu,\eta}\nmmuh&Z_{\nu,-\eta}\nmmuh\\
0&0&Z_{-\nu,\eta}\nmmuh&Z_{-\nu,-\eta}\nmmuh
\end{pmatrix},\\
W_\nu^{\mu,\rho}(n)&=
\begin{pmatrix}
Z^\dagger_{\mu,\rho}\npnuh&0&Z^\dagger_{\mu,-\rho}\npnuh&0\\
0&Z_{\mu,\rho}\nmnuh&0&Z_{\mu,-\rho}\nmnuh\\
Z^\dagger_{-\mu,\rho}\npnuh&0&Z^\dagger_{-\mu,-\rho}\npnuh\\
0&Z_{-\mu,\rho}\nmnuh&0&Z_{-\mu,-\rho}\nmnuh
\end{pmatrix}.
\end{split}
\end{equation}
We have seen that in $d\geq 3$, the fields $\Phi^{(\mu\nu)\dagger}$ and 
$\Phi^{(\rho\mu)}$ are also coupled, through a matrix $X_\mu^{\nu\rho}$; 
similarly, $\Psi^{(\mu\nu)\dagger}$ and $\Psi^{(\nu\eta)}$ are coupled 
through a matrix
$Y_\nu^{\mu\eta}$. As for $V$ and $W$, the lower index refers to the 
direction of the links carrying the elements of $Z$, $Z^\dagger$ which appear
in $X$ (or $Y$). These matrices have similar forms as the
matrices $V$, $W$ above (we won't need their explicit expression in the
following). 
These four sets of matrices can be grouped separately into matrices of 
size $N_b\times N_b$, which we call $\mathcal{V}(n)$, $\mathcal{W}(n)$, 
$\mathcal{X}(n)$, $\mathcal{Y}(n)$. These four matrices make up the complete 
coupling matrix $\mathcal{M}(n)$: the action reads
reads
\begin{equation}
-S_Z[n]=\begin{pmatrix}\Phi^{(\mu\nu)}\\
\Psi^{(\mu\nu)}\end{pmatrix}^\dagger  
\begin{pmatrix}\mathcal{X}&\mathcal{V}\\\mathcal{W}&\mathcal{Y}\end{pmatrix}
\begin{pmatrix}\Phi^{(\mu\nu)}\\
\Psi^{(\mu\nu)}\end{pmatrix}.
\end{equation}
Taking the mass term into account, the integral over the 
auxiliary fields yields 
\begin{equation}
\Det(m_b\mathbb{I}-\mathcal{M})^{-1}\propto 
\exp\left\{-\Tr\ln(1-m_b^{-1}\mathcal{M})\right\}
\approx \exp\left\{m_b^{-1}\Tr\mathcal{M}
+\frac{m_b^{-2}}{2}\Tr\mathcal{M}^2\right\},
\end{equation} 
where we performed the large-$m_b$ expansion up to second order.
To analyze the traces, we use the `selection rules' given by 
\eqref{selection}. 
$X_\mu^{\nu\rho}$ connects planes 
$(\mu\nu)>(\rho\mu)$, therefore its block appears under the diagonal 
in the matrix $\mathcal{X}$; on the opposite, $Y_\nu^{\mu\eta}$ 
connects planes $(\mu\nu)<(\nu\eta)$, so its block is over the
diagonal in $\mathcal{Y}$. As a result, $\Tr\mathcal{M}=0$, and 
$\Tr\mathcal{X}^2=\Tr\mathcal{Y}^2=0$. Therefore, the first nontrivial term
appears at the order $1/m_b^2$, and takes the value
$\Tr\mathcal{M}^2=2\Tr(\mathcal{VW})$.

To compute this term, we notice
that the block $V_\mu^{\nu,\eta}$ connects planes $(\mu\nu)$, $(\mu\eta)$ 
sharing the
same lower index $\mu$ (that is, positive generators $e_{\mu\nu}$, 
$e_{\mu\eta}$
situated on the same row); on the opposite, a block $W_\nu^{\mu,\rho}$
connects generators situated on the same column. Therefore, through
$\mathcal{V}$ we can jump along a row, and through $\mathcal{W}$ we jump along
a column. When computing $2\Tr(\mathcal{VW})$ we want to be back at the 
initial position after two jumps, so that only `immobile jumps' are 
allowed:
\begin{equation}
\frac{1}{2}\Tr\mathcal{M}^2=
\Tr(\mathcal{VW})=\sum_{\mu<\nu}\Tr(V_\mu^{\nu,\nu}W_\nu^{\mu,\mu}).
\end{equation}
Thus, to this order the planes are ``decoupled'' from one another,
and the contribution of each plane is identical to what we had 
found in the two-dimensional framework (see Eq.~\eqref{trace-d2}):
\begin{multline}\label{trace-d}
\Tr\left(\mathcal{V}(n)\mathcal{W}(n)\right)=
\sum_{1\leq\mu<\nu\leq d}Z^\dagger_{\nu,\nu}\npmuh Z^\dagger_{\mu,\mu}\npnuh
+ Z^\dagger_{-\nu,-\nu}\npmuh Z_{\mu,\mu}\nmnuh\\ 
+ Z_{\nu,\nu}\nmmuh Z^\dagger_{-\mu,-\mu}\npnuh 
+  Z_{-\nu,-\nu}\nmmuh Z_{-\mu,-\mu}\nmnuh.
\end{multline}
Thus, as in two and three dimensions,
only the diagonal elements of the $Z$-fields contribute to 
the action up to  order $1/m_b^{2}$. Each of the above terms 
is the product of two diagonal matrix elements 
which self-coupled auxiliary fields carried by the same plaquette, so
each term can be associated with a well-defined plaquette (we come back
to this property in next section).
The higher-order terms in $1/m_b$ are more complicated, since 
the matrices $X$, $Y$ and 
non-diagonal blocks $V$, $W$ start contributing.

To summarize, the effective action to second order in $1/m_b$ 
has the same structure in any dimension:
\begin{equation}       \label{cft-bar-z-d}
-S[Z]  = \sum_n  
m_b^{-2} \sum_{\mu<\nu}\Tr(V_\mu^{\nu,\nu}W_\nu^{\mu,\mu})(n) + 
\sum_{\alpha=1}^d \Tr \ln \left(1 -  Z\npalh Z^\dagger\npalh\right) .
\end{equation}


\subsection{Stationary point of the large-$m_b$ effective action}

The factor $N_c$ in front of the action $S[Z]$ suggests
to study the large-$N_c$ limit of the theory, that is look for stationary
points
of this action with respect to variations of the matrix fields $Z$, 
$Z^\dagger$. Since
we computed the action $S[Z]$ up to second order in $1/m_b$, we will
keep this approximation \eqref{cft-bar-z-d} and compute its 
saddle-point equations. 

The variation of the quadratic terms $\Tr(\mathcal{VW})$ is easy to compute
from \eqref{trace-d}:
it only involves variations of diagonal elements of the matrices $Z$ or 
$Z^\dagger$. On the opposite, the variation of the second term in 
Eq.~\eqref{cft-bar-z-d} involves all matrix elements:
\begin{equation}
\begin{split}
\delta\Tr\ln(1 -  ZZ^\dagger)&=-\Tr\left[
\delta Z\ Z^\dagger(1-ZZ^\dagger)^{-1}
+\delta Z^\dagger\ Z(1-Z^\dagger Z)^{-1}\right]\\
&=-\sum_{a,b}\delta Z_{ab}\left(Z^\dagger(1-ZZ^\dagger)^{-1}\right)_{ba}
+\delta Z^\dagger_{ab}\left(Z(1-Z^\dagger Z)^{-1}\right)_{ba}.
\end{split}
\end{equation}
Therefore, setting $\frac{\delta S[Z]}{\delta Z_{ab}}=0$ for all $a\neq	b$
implies that the matrix $Z(1-Z^\dagger Z)^{-1}$ is diagonal;
this implies that $Z$ is itself a diagonal matrix. We then compute
the saddle point equations with respect to 
variations of the diagonal elements $Z_{aa}$.
For any site $n$ and $\mu\neq\nu$, Eq.~\eqref{trace-d} yields the following variations:
\begin{equation}\label{saddle}
\frac{\delta S[Z]}{\delta Z^\dagger_{\nu,\nu}\npmuh}
= \frac{Z_{\nu,\nu}\npmuh}{1-|Z_{\nu,\nu}\npmuh|^2}
-m_b^{-2}Z^\dagger_{\mu,\mu}\npnuh,
\end{equation}
and similar expressions for the variation of $S[Z]$ with respect to the 
components
$Z^\dagger_{-\nu,-\nu}\npmuh$, $Z_{\nu,\nu}\nmmuh$ and $Z_{-\nu,-\nu}\nmmuh$. 
Setting these variations to zero, we get the full set of saddle-points
equations. These equations obviously admit the trivial configuration 
$Z\equiv 0$ as solution. Taking into account the 
condition $m_b\gg 1$, one can show that this solution is the unique
one.

It therefore makes sense to expand the action \eqref{cft-bar-z-d} 
to quadratic order in $Z$, $Z^\dagger$:
\begin{equation}\label{action-quad}
-S[Z]_{\rm quad}  =m_b^{-2}  \sum_n  \left(
\sum_{\mu<\nu}\Tr(V_\mu^{\nu,\nu}W_\nu^{\mu,\mu})(n) - 
\sum_{\alpha=1}^d \sum_{a,b=1}^{d'}m_b^2\,|Z_{ab}\npalh|^2\right).
\end{equation}
From the expression \eqref{trace-d}, the above action seems to describe
free bosonic fields. The nondiagonal elements $Z_{ab}$ with $a\neq b$ 
are nondynamical, since they only appear in the mass term. On the opposite,
the diagonal terms $Z_{aa}$ appear both in the mass term and the 'kinetic
energy term' \eqref{trace-d}, so they seem to correspond to propagating modes. This is 
actually not the case: as we already noticed, the  terms 
Eq.~\eqref{trace-d} only couple matrix elements related to the same
plaquette, so that these fields can only propagate around
one plaquette. 
The diagonal fields are
therefore non-propagating modes as well, so the above quadratic action 
is non-dynamical. This is
not so surprising, since the auxiliary bosonic fields $\phi(n)$ were
from the beginning also confined to one plaquette.
The same phenomenon persists if one includes several generations
$n_b>1$.

Propagation can be induced by including higher-order terms in
$ZZ^\dagger$ when expanding the logarithm. This way, one obtains a
quartic contribution $\Tr(ZZ^\dagger)^2$, which allows to couple 
together different diagonal elements through non-diagonal ones. This 
contribution includes for instance terms of the form 
$
Z_{2,2}Z^\dagger_{2,-2}Z_{-2,-2} Z^\dagger_{-2,2}
$ (all elements on the link $(n+\hat 1/2)$),
which couple fields carried by two adjacent plaquettes, namely the
plaquette $(n,1,2)$ carrying $Z_{2,2}\npoh$, and the plaquette 
$(n,1,-2)$ carrying $Z_{-2,-2}\npoh$ (see Fig.~\eqref{dual-2d}).

\section{Concluding remarks}
We have considered the dual formulation of the lattice theory which induces 
Wilson's pure gauge action after integrating over auxiliary bosonic 
fields, in the limit of large mass and many ``generations''.
In our model the ``flavor'' 
degrees of freedom are associated not only with the number of the ``generations'' of 
the inducing field, but also with a particular
plaquette; besides, we need fields with left. resp.. right ``chirality'',
which doubles the number of flavor degrees of freedom.

We investigate the properties 
of the lattice model in the simpler case of one generation.
The structure of the inducing theory allows us to apply the 
color-flavor transformation to obtain a `dual' effective theory in terms 
of colorless matrices $Z$ carried by the lattice links. After integrating 
over the auxiliary bosons, we obtain an effective action uniquely in terms
of the $Z$ fields, which is computed explicitly in the limit of large
auxiliary mass, leading to a trivial non-propagating theory in the
large-$N_c$ limit.

The color-flavor transformation for the $SU(N_c)$ gauge group 
yields some differences, 
related with the decomposition of the colorless sector 
into disconnected subsectors labeled by the baryonic charge $Q$ 
\cite{BNSZ,su-rivals}. 
Our above derivations
correspond to the sectors $Q=0$ with no contribution of closed baryon loops 
\cite{SchWett}. The investigation of the effect of these loops is in progress.
Note that the choice of  $U(N_c)$ instead of the realistic $SU(N_c)$ is 
irrelevant in the large-$N_c$ limit.

\medskip
\paragraph{Acknowledgements}

This research is inspired by numerous 
discussions with S.~Nonnenmacher, who suggested a general strategy of the  
integration over the auxiliary fields in $d$-dimensions \cite{ShNonn}, and with
J.~Budczies, who independently constructed $d=2$ 
one-plaquette effective action of the Z-field \cite{J-diss}. 
I am grateful to D.~Diakonov,  V.~Petrov, M.~Polyakov and 
M.R.~Zirnbauer for useful discussions and comments. 
I would like to acknowledge the hospitality at the Abdus Salam 
International Center for Theoretical Physics where this work was completed.


\begin{thebibliography}{abcdef}

\bibitem{Wilson74}K.~Wilson, Phys.\ Rev. {\bf D 10} (1974) 2445.

\bibitem{Gross80}D.~Gross and E.~Witten,  Phys.\ Rev. {\bf D 21} (1980) 446.


\bibitem{Banks77}T.~Banks, R.~Myerson and J.~Kogut, 
Nucl.\ Phys. {\bf B129} (1977) 493.

\bibitem{Peskin78} M.E.~Peskin,  Ann.\ Phys. {\bf 113} (1978) 122.

\bibitem{Polley91}L.~Polley and U.J.~Wiese,  
Nucl.\ Phys. {\bf B356} (1991) 629.

\bibitem{Halpern79}M.B.~Halpern,  Phys. Rev. {\bf D19} (1979) 517.

\bibitem{Rusakov97}B.~Rusakov,  Phys. Lett. {\bf B398} (1997) 331.

\bibitem{Anishety93}R.~Anishety, S.~Cheluvaraja and H.S.~Sharatchandra, 
Phys.\ Lett. {\bf B314} (1993) 387.

\bibitem{DP2000}D.~Diakonov and V.~Petrov, hep-th/9912268.

\bibitem{Zirn}M.R.~Zirnbauer, J. Phys. {\bf A29} (1996) 7113; {\it ibid.}, 
Contribution to the Proceedings of the XIIth International Congress of Mathematical Physics, Brisbane, 1997.

\bibitem{Bander}M.~Bander,  Phys.\ Lett.\ {\bf B126} (1983) 463.

\bibitem{Hamber}H.~Hamber.  Phys.\ Lett.\ {\bf B126} (1983) 471.

\bibitem{Kogan92}I.I.~Kogan, G.W.~Semenoff and N.~Weiss, Phys.\ Rev.\ Lett.
{\bf 69} (1992) 3435.

\bibitem{Kazakov93}V.A.~Kazakov and A.A.~Migdal, 
Nucl.\ Phys. {\bf B397} (1993) 214.

\bibitem{Makeenko93}Yu.~Makeenko and K.~Zarembo,  Nucl.\ Phys. 
{\bf B422} (1994) 237.

\bibitem{Hasen92}A.~Hasenfratz and P.~Hasenfratz, Phys.\ Lett. 
{\bf B297} (1992) 166.  

\bibitem{Aref}I.Ya.~Aref'eva, Phys.\ Lett.\ {\bf B308} (1993) 347. 

\bibitem{Zarembo}K.~Zarembo, Phys.\ Dokl.\ {\bf 38} (1993) 236. 

\bibitem{BNSZ}  J.~Budczies, S.~Nonnenmacher, Ya.~Shnir and M.R.~Zirnbauer, 
Nucl.\ Phys. {\bf B635} (2002) 309.
\bibitem{BuZirn} J.~Budczies and M.R.~Zirnbauer, {\it Howe duality 
for an Induced Model of Lattice U(N) Yang-Mills Theory}, ~math-ph/0305058.


\bibitem{Suranyi}P.~Suranyi, Phys.\ Rev. {\bf D57} (1998) 5084.

\bibitem{su-rivals} B.~Schlittgen and T.~Wettig, Nucl.\ Phys. {\bf B632} (2002) 155.

\bibitem{SchWett}B.~Schlittgen and T.~Wettig,  Nucl.\ Phys.\ Proc.\ Suppl. {\bf 119} 
(2003) 956;  ~hep-th/0208044.
\bibitem{SchWei}Yi Wei and T.~Wettig,  {\it Bosonic Color-Flavor 
Transformation for the Special Unitary Group},~hep-lat/0411038 


\bibitem{J-diss} J.~Budczies, {\it The Color-Flavor Transformation and its application to 
Quantum Chromodynamics}, PhD Thesis, University of Cologne, 2002.
\bibitem{ShNonn}  S.~Nonnenmacher and Ya.~Shnir {\it  The Color-Flavor Transformation of induced 
QCD}, ~hep-lat/0210002.
\bibitem{ictp}Ya.~Shnir, ICTP preprint IC/2002/124, Trieste, 2002. 
\end{thebibliography}
\end{document}